\documentclass[12pt,english]{iopart}
\usepackage{iopams}
\usepackage{babel}
\usepackage{graphicx}
\usepackage[T1]{fontenc}
\usepackage[latin9]{inputenc}
\begin{document}

\title[]{Magnetic domain walls induced by twin boundaries in low doped Fe-pnictides}

\author{Bo Li, Jian Li, Kevin E Bassler and C S Ting}

\address{Department of Physics and Texas Center for
Superconductivity, University of Houston, Houston, Texas 77204, USA}
\begin{abstract}
Inspired by experimental observations of the enhancement of
superconductivity at the twin-boundary (TB) in slightly electron
doped Ba(Ca)(FeAs)$_2$ where a strong $2\times1$ antiferromagnetic
(AF) collinear order is in presence, we investigate theoretically
the effects of TBs on the complex interplay between magnetism and
superconductivity using a minimum phenomenological two-orbital
model. The magnetic order can be simulated from an onsite Coulomb
interaction $U$ and the Hund's coupling $J_H$, while the effective
pairing interaction yields the superconductivity with $s\pm$ wave
symmetry. The spatial distributions of the magnetic, superconducting
and charge density orders near two different types of TBs are
calculated. Each of the TBs has two different orientations. We find
that the first type TBs, which corresponds to a $90^\circ$ lattice
rotation in the $a-b$ plane, enable magnetic domain walls (DWs) to
be pinned at them, and that superconductivity is enhanced at such
TBs or DWs. This result is consistent with experiments for a TB with
an orientation of $45^\circ$ from the x-axis. Contrastingly, we
predict that superconductivity is suppressed at the second type of
TBs which correspond to an asymmetrical placement of As atoms on the
opposite sides of the TB. Furthermore, the lattice-mismatch effect
across the TBs is investigated. The comparison of our results with
the observations from the nuclear-magnetic-resonance (NMR)
experiments are also discussed.
\end{abstract}

\maketitle

\section{Introduction}
The recent discovery of Fe-pnictide based superconductors offers an
alternative avenue to explore the physics of a new family of high
temperature superconductors \cite{Kamihara,Ren,Chen,Cruz,Chen08}.
Similar to the cuprates, the parent compounds of the FeAs-based
superconductors also possess antiferromagntic (AF) ground states
\cite{Cruz,Chen08}. With increasing electron or hole doping, the AF
order is suppressed and superconductivity (SC) appears in both the
cuprates and the Fe-pnictides. However, different from the cuprates,
SC and a $2\times1$ collinear AF or spin-density-wave (SDW) order
can coexist in doped Ba(FeAs)$_2$ superconductors
\cite{Laplace,Julien}. Because each unit cell of these new materials
contains two inequivalent Fe ions, different organizations of the
magnetic moments of Fe ions in both normal and superconducting
states can lead to a diverse assortment of magnetic structures and
unusual electronic properties \cite{Mazin,Gorkov}.

Recently, twin boundaries (TBs) oriented $45^\circ$ from the
x(a)-axis were observed in the normal state of
Ca(Fe$_{1-x}$Co$_{x}$As)$_{2}$ with $x\sim0$ by scanning tunneling
microscopy (STM) experiments \cite{Chuang}. Across these TBs, the
a-axis of the crystal rotates by $90^\circ$, and the modulation
direction of AF order that exists is rotated by $90^\circ$ as well.
That is, $90^\circ$ magnetic domain walls (DWs) are formed at the
TBs.  Also, in the SC state of underdoped
Ba(Fe$_{1-x}$Co$_{x}$As)$_{2}$ with $x<0.07$, it has been found that
the diamagnetic susceptibility is increased and that the superfluid
density is enhanced on the same type of TB in
superconducting-quantum-interference-device miscroscopy (SQIDM)
experiments \cite{Kalisky}. Consistent with these experiments, a
theoretical study \cite{Huang} found that $90^\circ$ DWs can be
formed at low doping levels and that SC is enhanced on them.
However, the DWs considered in that study were formed in the absence
of TBs, and were induced instead by a strong effective Coulomb
interaction between charge carriers, while in the experiments
\cite{Chuang,Kalisky} the DWs were pinned at TBs.

In this work, in order to better understand the effects that TBs
have on the spatial profiles of magnetic and superconducting order,
we investigate the magnetic, SC and charge density orders near two
different types of TBs using the Bogoliubov-de-Gennes (BdG)
equations for very under-doped Ca(or Ba)(FeAs)$_{2}$ compounds. The
lattices on the opposite sides of the type-1 TB have a $90^\circ$
orientation difference, and the As atoms across the type-2 TB are
asymmetrically placed. The present approach is based upon a
two-orbital model \cite{Zhang}. This model takes into account of the
asymmetry of the As atoms below and above of the Fe plane which
should be suitable for describing the cleaved surface layer or
layers in surface sensitive experiments. As a result, the obtained
phase diagram for the electron doped Ba(Fe$_{1-x}$Co$_{x}$As)$_{2}$
\cite{Zhou} is in qualitative agreement with experiments
\cite{Laplace,Julien,Wang,Pratt,Christianson}. The obtained Fermi
surface evolution as a function of doping \cite{Zhou} and the Fermi
surface at zero doping \cite{Zhou11} are consistent with angular
resolved photo-emission spectroscopy (ARPES) experiments
\cite{Terashima, Sekiba, Richard} and the electron-hole Dirac cone
states as observed by magnetoresitance experiments \cite{Huynh}.
Another critical test of the model is that the resonance peak
observed at negative energy in a vortex core by STM experiments
\cite {Shan} could only be explained by the present model
\cite{Gao}. The approach has also been used to study the superfluid
density \cite{Huang1} as the doping and temperature varying, and the
results are in good agreement with experiments in films \cite{Yong}.
We predict that the enhancement or suppression of SC, the location
of DWs and the electron-density distributions are largely dependent
on the nature of TBs. The relationship between magnetism, SC and the
charge order is always one of the most important problems in the
research of unconventional superconductors. Our work provides the
first study on different kinds of twin boundaries, which involves
all the three aspects mentioned above. More importantly, it shows
where and how the superconductivity can be enhanced/suppressed in
the presence of TBs.

This paper is constructed as follows. In Section 2, brief
descriptions of the model Hamiltonian and methodology will be given.
In Section 3, we calculate the phase diagram according to the
interaction parameters used in the present work. In Section 4, the
numerical results on the spatial structures of magnetic,
superconducing and charge orders for two types of TBs are presented
in the very under-doped region (x=0.04) where the SC order is weak
and the magnetic order is strong. Each of the TBs has two
configurations, which either aligns along $90^\circ$ or $45^\circ$
from the x-axis. In Section 5, we approximate the lattice mismatch
across one of the TBs by a disordered potential along the TB, and
discuss how the magnetic, SC and charge orders near the TB are
affected by the strength of the disorder. In Section 6, we compare
our results with the magnetic DWs observed by nuclear magnetic
resonance experiments (NMR) \cite{Xiao}. Finally concluding remarks
are made in the final section.

\section{Hamiltonian and Methodology}
Consider the Hamiltonian $H=H_0+H_{SC}+H_{int}$ that describes the
energy of charge carriers. $H_0$ is a non-interacting two-orbital
tight-banding model from \cite{Zhang,Zhou},
\begin{eqnarray}
H_{0}=-\sum_{{i}\mu{j}\nu\sigma}(t_{{i}\mu{j}\nu}c^{\dagger}_{{i}\mu\sigma}
c_{{j}\nu\sigma}+\textrm{H.c.})
-t_{0}\sum_{{i}\mu\sigma}c^{\dagger}_{{i}\mu\sigma}c_{{i}\mu\sigma}
\end{eqnarray}
where $t_{{i}\mu{j}\nu}$ is the hopping parameter between two
electrons, one at position ${i}$ with the orbital $\mu$ and the
other at position ${j}$ with orbital $\nu$, and
$c^{\dagger}_{{i}\mu\sigma}$ is the creation operator of an electron
with spin $\sigma$ at position ${i}$ with orbital $\mu$. $t_0$ is
the chemical potential. The pairing interaction energy of the
electrons is
\begin{eqnarray}
H_{SC}=\sum_{{i}\mu{j}\nu\sigma}(\Delta_{{i}\mu{j}\nu}c^{\dagger}_{{i}\mu\sigma}
c^{\dagger}_{{j}\nu\bar{\sigma}}+\textrm{H.c.})
\end{eqnarray}
where $\Delta_{{i}\mu{j}\nu}$ is the pairing parameter between two
electrons. Here $\bar{\sigma}$ denotes the opposite spin of
$\sigma$. The mean-field magnetic interaction energy \cite{Zhou} is
\begin{eqnarray}
\fl \nonumber
H_{int}=(U-3J_H)\sum_{{i},\mu\ne\nu,\sigma}\langle{n_{{i}\mu\sigma}}\rangle{n_{{i}\nu\sigma}}
+(U-2J_H)
\sum_{{i},\mu\ne\nu,\sigma\ne\bar{\sigma}}\langle{n_{{i}\mu\bar{\sigma}}}\rangle{n_{{i}\nu\sigma}}
\\
+U\sum_{{i},\mu,\sigma\ne\bar{\sigma}}\langle{n_{{i}\mu\bar{\sigma}}}\rangle{n_{{i}\mu\sigma}}
\end{eqnarray}
where $U$ is the on-site Coulomb interaction, $J_H$ is the Hund's
coupling, $n_{{i}\mu\sigma}$ is the electron number operator, and
$\langle n_{{i}\mu\sigma} \rangle$ is the local electron density.
The eigenvalues and eigenfunctions of the total Hamiltonian $H$ can
be obtained by self-consistently solving the BdG equations
\begin{eqnarray}
\sum_{{j},\nu}\left( \begin{array}{ccc}
H_{{i}\mu{j}\nu\sigma}&\Delta_{{i}\mu{j}\nu}
\\ \Delta^{*}_{{i}\mu{j}\nu}&-H^{*}_{{i}\mu{j}\nu\bar{\sigma}}
\end{array} \right)  \left( \begin{array}{ccc}
u^{n}_{{j}\nu\sigma} \\ v^{n}_{{j}\nu\bar{\sigma}}
\end{array} \right)=E_n\left(
\begin{array}{ccc}u^{n}_{{i}\mu\sigma} \\
v^{n}_{{i}\mu\bar{\sigma}} \end{array} \right)
\end{eqnarray}
where
\begin{eqnarray}
\fl
H_{{i}\mu{j}\nu\sigma}=-t_{{i}\mu{j}\nu}+[U\langle
n_{{i}\mu\bar{\sigma}}\rangle +(U-2J_H)\langle
n_{{i}\bar{\mu}\bar{\sigma}}\rangle
+(U-3J_H)\langle
n_{{i}\bar{\mu}\sigma}\rangle-t_0]\delta_{{i}{j}}\delta_{\mu\nu}
\end{eqnarray}
is the matrix element of $H$ with the same spin $\sigma$ between the
orbital $\mu$ at position ${i}$ and the orbital $\nu$ at position
${j}$, and $t_0$ is the chemical potential. The pairing parameter
$\Delta_{{i}\mu{j}\nu}$ and the local electron densities $\langle
n_{{i}\mu\uparrow}\rangle$ and $\langle n_{{i}\mu\downarrow}\rangle$
satisfy the following self-consistent conditions
\begin{eqnarray}
\Delta_{{i}\mu{j}\nu} & = &
\frac{V_{{i}\mu{j}\nu}}{4}\sum_{n}(u^{n}_{{i}\mu\uparrow}
v^{n*}_{{j}\nu\downarrow} +u^{n}_{{j}\nu\uparrow}
v^{n*}_{{i}\mu\downarrow})\tanh\left(\frac{E_n}{2k_BT}\right)
\\
\langle n_{{i}\mu\uparrow}\rangle & = &
\sum_{n}\left|u^{n}_{{i}\mu\uparrow}\right|^{2}f(E_n)
\\
\langle n_{{i}\mu\downarrow}\rangle & = &
\sum_{n}\left|v^{n}_{{i}\mu\downarrow}\right|^{2}[1-f(E_n)]
\end{eqnarray}
where $V_{{i}\mu{j}\nu}$ is the pairing interaction between next
nearest neighboring (NNN) sites i and j which gives rise to the
superconductivity with $s\pm$ wave symmetry \cite{Zhang, Zhou}, and
$f(x)$ is the Fermi-Dirac distribution function. The SC order
parameter at position ${i}$ is defined as $
\Delta_{{i}}\equiv\left|\Delta_{{i},{i}+\hat{x}+\hat{y}}+\Delta_{{i},{i}-\hat{x}-\hat{y}}+
\Delta_{{i},{i}+\hat{x}-\hat{y}}+\Delta_{{i},{i}-\hat{x}+\hat{y}}\right|$,
the local magnetic moment at position ${i}$ is defined as $
m_{{i}}\equiv\frac{1}{4}\sum_{\mu}(\langle
n_{{i}\mu\uparrow}\rangle-\langle n_{{i}\mu\downarrow}\rangle) $,
and the total charge density at position ${i}$ is defined as $
\langle n_{{i}}\rangle \equiv\sum_{\mu}(\langle
n_{{i}\mu\uparrow}\rangle+\langle n_{{i}\mu\downarrow}\rangle) $.
The chemical potential $t_0$ is determined by the electron filling
per site $n$ ($n=2+x$), and for the value of the hopping terms
$t_{{i}\mu{j}\nu}$ are assumed to be $t_{1-4}=1,0.4,-2,0.04$
\cite{Zhang,Zhou}. Only the electron pairings of the same orbital
between the next-nearest-neighbor Fe sites are considered. For
example, we choose $V_{{i}\mu{j}\nu}=1.4$ for $\mu=\nu$ and
$\left|{i}-{j}\right|=\sqrt{2}$, and zero for all other cases. This
choice of the pairing potential gives rise to the SC order with
$s_{\pm}$-wave symmetry \cite{Seo,Mazin08}.

\section{Phase Diagram}

\begin{figure}[t]
\centering
\includegraphics[width=2.5in] {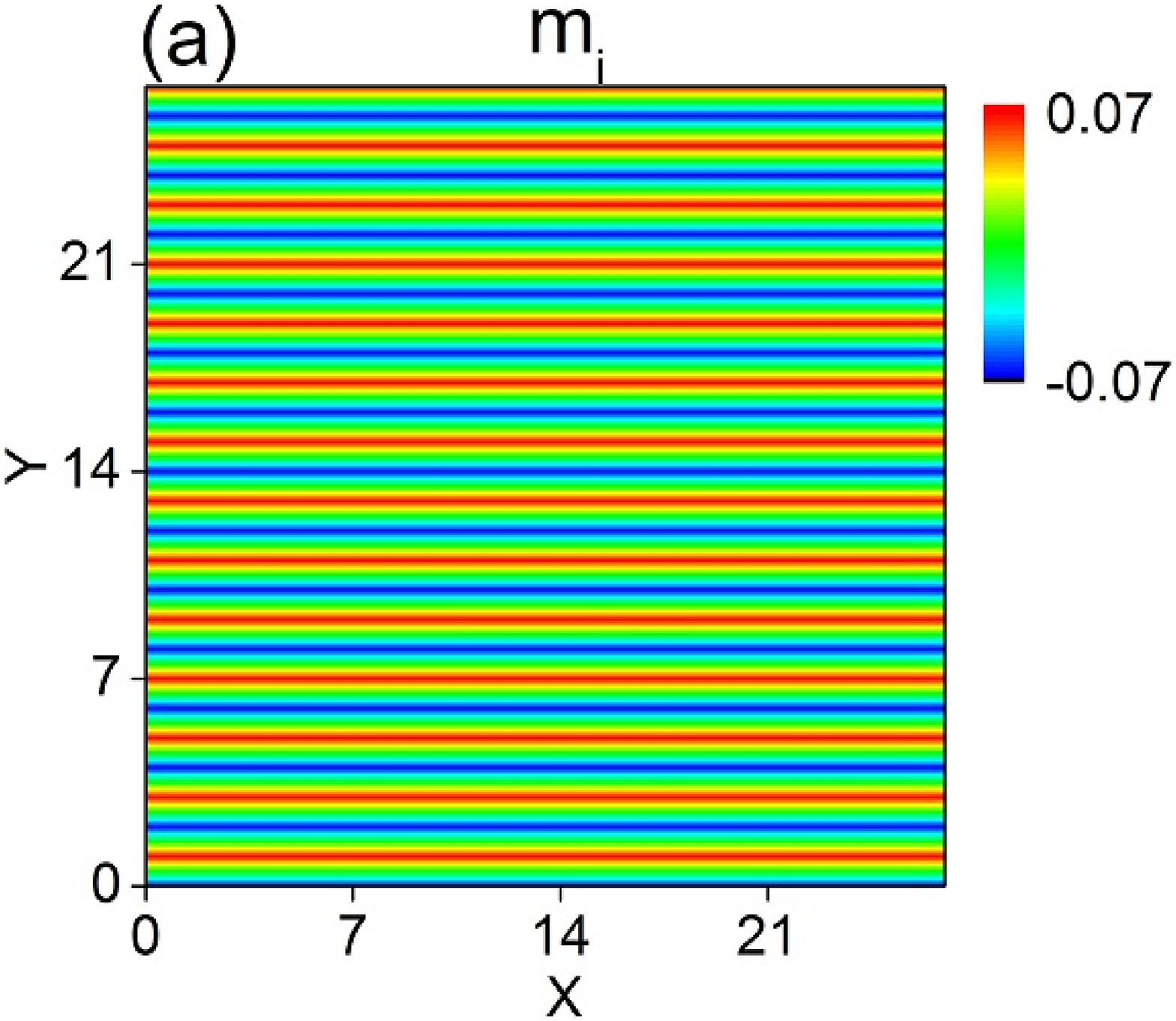}
\includegraphics[width=2.5in] {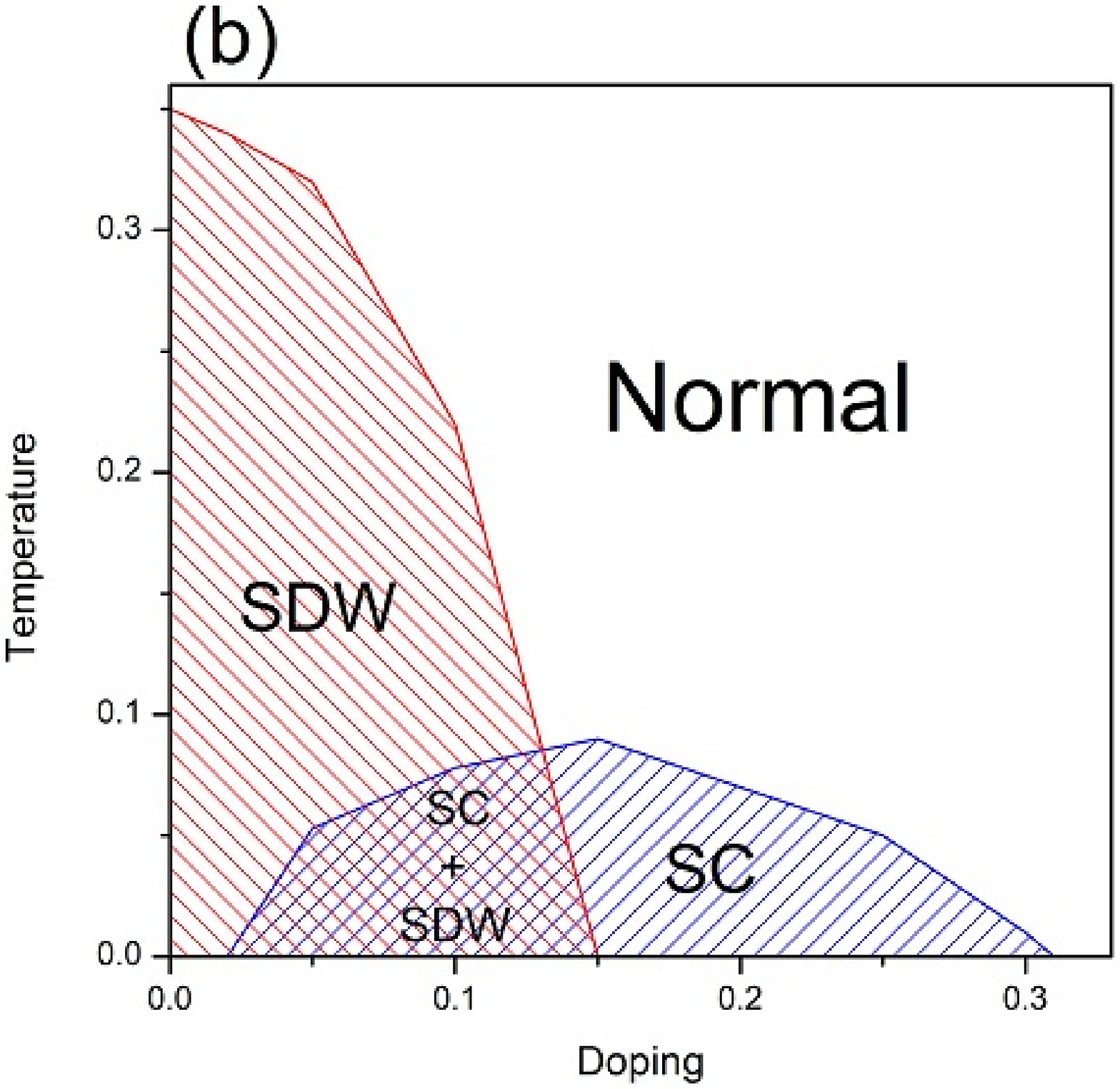}
\caption{($a$) The orientations of magnetic moments of Fe ions on a
28x28 lattice at zero temperature and doping $x=0.04$. ($b$) The
phase diagram as a function of electron doping (x).} \label{phase}
\end{figure}

The phase diagram of the electron-doped Ba(Fe$_{1-x}$Co$_x$As)$_2$
compounds as a function of temperature $T$ and doping $x$ has been
qualitatively mapped out with $U=3.4$, $J_H=1.3$, and the NNN
pairing interaction $V=1.2$ \cite{Zhou}. The distribution of the
$2\times1$ collinear AF order and the SC order are spatially
uniform. In Reality,the lattice in the Fe-plane of these compounds
is almost square, having slightly different lattice constants $a$
and $b$ ($a>b$) along $x$- and $y$-directions \cite{Kim}. In the
present work, nearest-neighbor hoping terms are chosen to be
$t_a=1.0$ and $t_b=1.2$,  we also slightly increase the magnitudes
of our interaction parameters ($U=3.8$, $J_H=1.3$, and $V=1.4$) in
order to generate more pronounced inhomogeniety of the order
parameters near the TBs. With the new set of the interaction
parameters, we recalculate the magnetic configuration at $x=0.4$ on
a $28\times28$ lattice and the phase diagram as a function of the
electron doping which are respectively shown in figures
\ref{phase}($a$) and ($b$). Figure \ref{phase}($a$) exhibits the
$2\times1$ collinear AF order, where the spins of the Fe ions on the
red lines are pointing upward, and the spins on the blue lines are
pointing downward. Since the magnetic order is originated in
itinerant interacting electrons, the magnetic configuration in
figure \ref{phase}($a$) is also refered as the SDW order. The spins
of the Fe ions are always  aligned ferromagnetically along a (or
x)-direction. The obtained phase diagram indicates that the SC order
is completely suppressed by the SDW near $x=0$. The SDW and SC are
coexisting with each other between $x\sim0.02$ and $x\sim0.15$. For
$x>0.15$ while the SDW order disappears, the SC still prevails. All
these features in our phase diagram are in qualitative agreements
with both surface- and bulk-sensitive experiments \cite{Laplace,
Julien, Wang, Pratt, Christianson}.

\section{Magnetic Domain-Wall Structures and Twin-Boundaries}
In the absence of the TBs, the SDW order discussed above is unstable
against the formation of the $90^\circ$ magnetic DWs oriented
$45^\circ$ from the x-axis as the strength of $U$ is increased to
$U=4.8$ for small $x$ at $T=0$ \cite{Huang}. In the presence of TBs,
the magnetic DWs could be formed at much weaker $U$ as demonstrated
in the present work. There are two types of TBs. The first one is
that the orientations of the lattices across the TB differ by a
$90^\circ$ rotation. The second one is that the As atoms are
asymmetrically placed across the TB as compared with a perfect
crystal lattice. For each type of the TBs, there are two
configurations: one is oriented along $90^\circ$ from x-axis and the
other is oriented along $45^\circ$ from the x-axis. In the
following, the spatial profiles of magnetic, SC and charge density
order near four different TBs at $T=0$ are investigated. Throughout
this work, we set $x=0.04$, $U=3.8$ and $J_H=1.3$ and $V=1.4$. Note
that DWs do not form spontaneously in the absence of TBs at these
parameter values, and all the order parameters have uniform
solutions.

\subsection{A. Type-1 Twin-Boundary Oriented $45^\circ$ From The x-Axis}
\begin{figure}[t]
\centering
\includegraphics[width=2.5in] {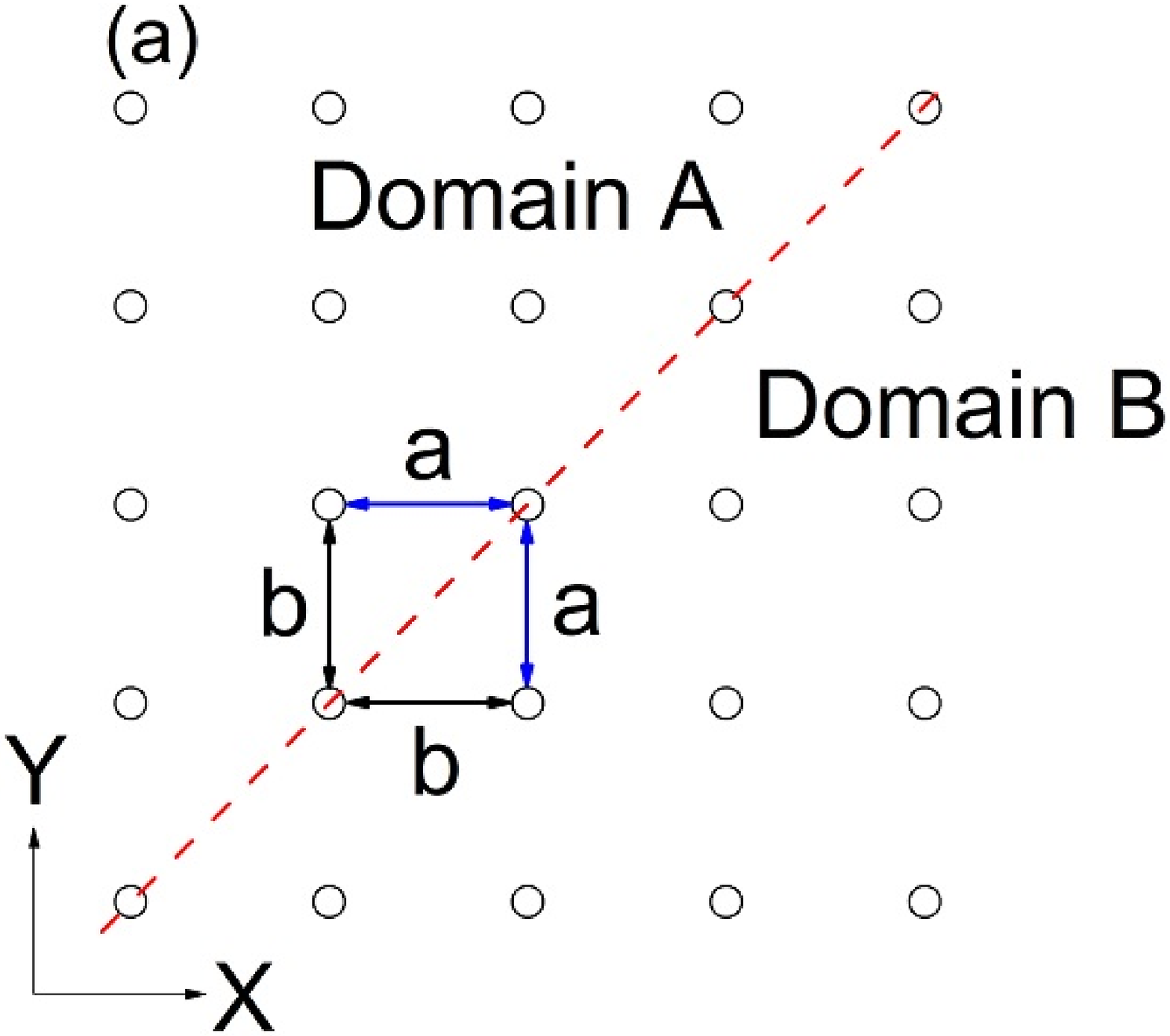}
\includegraphics[width=2.5in] {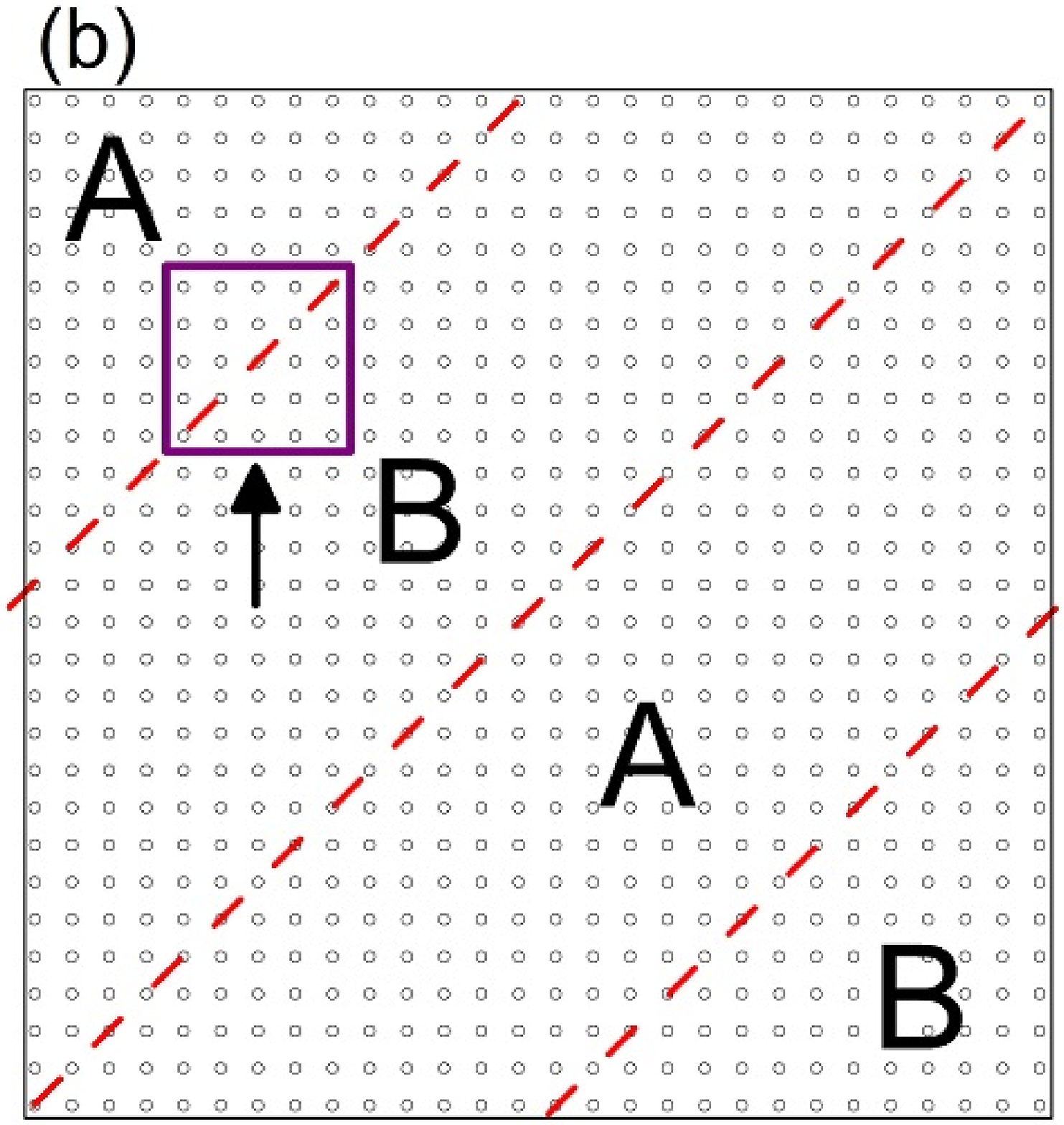}
\caption{($a$) The lattice structure near a diagonal TB (red dashed
line), the open circles represent the positions of Fe atoms, a (blue
solid line) and b (black solid line) are the lattice constants along
x and y directions in domain A. ($b$) Three TBs in the $28\times28$
lattice, the black arrow indicates the position of the lattice
structure shown in ($a$) in the whole lattice.}\label{Figa0}
\end{figure}

\begin{figure}[t]
\centering
\includegraphics[width=2.0in] {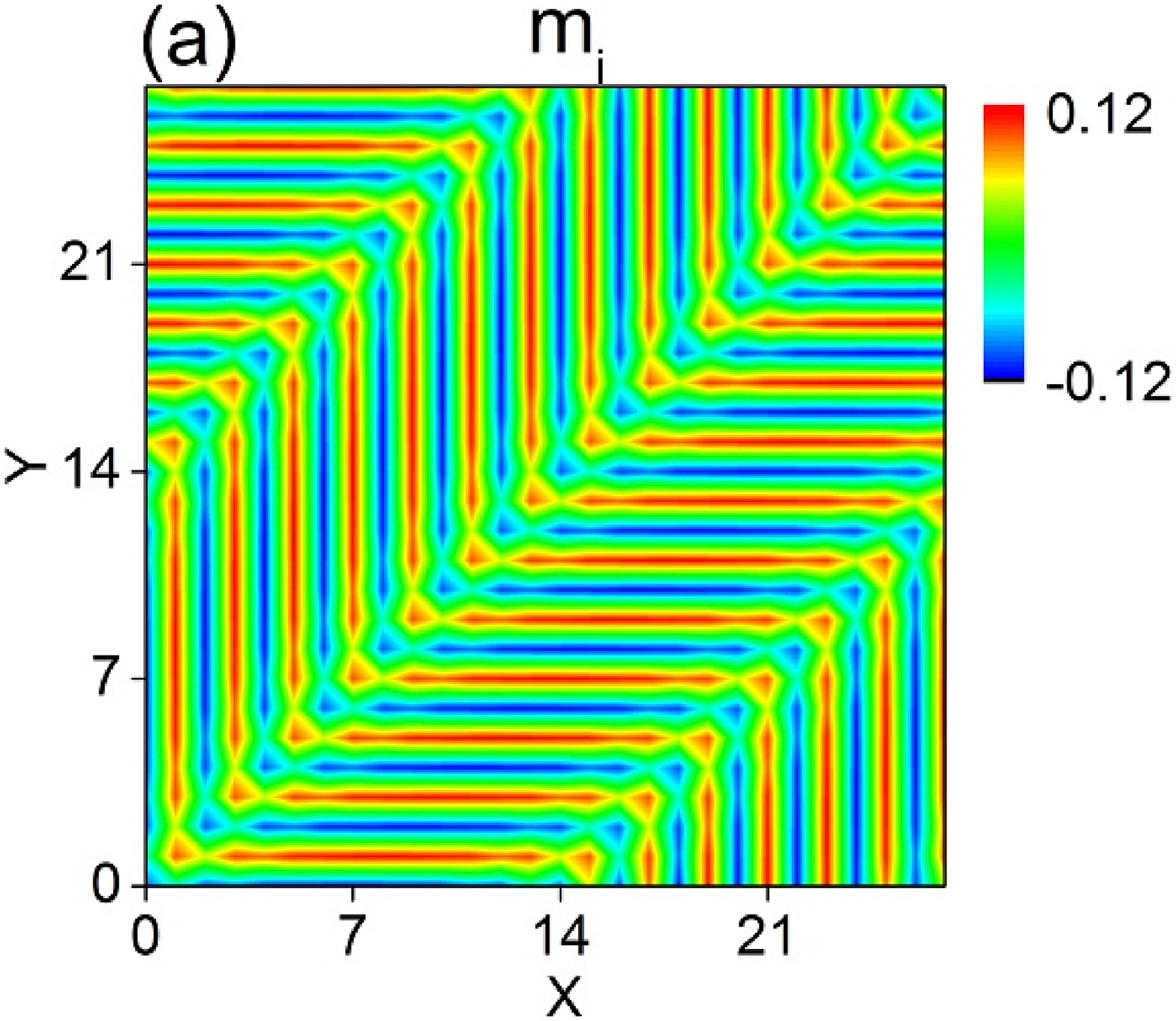}
\includegraphics[width=2.0in] {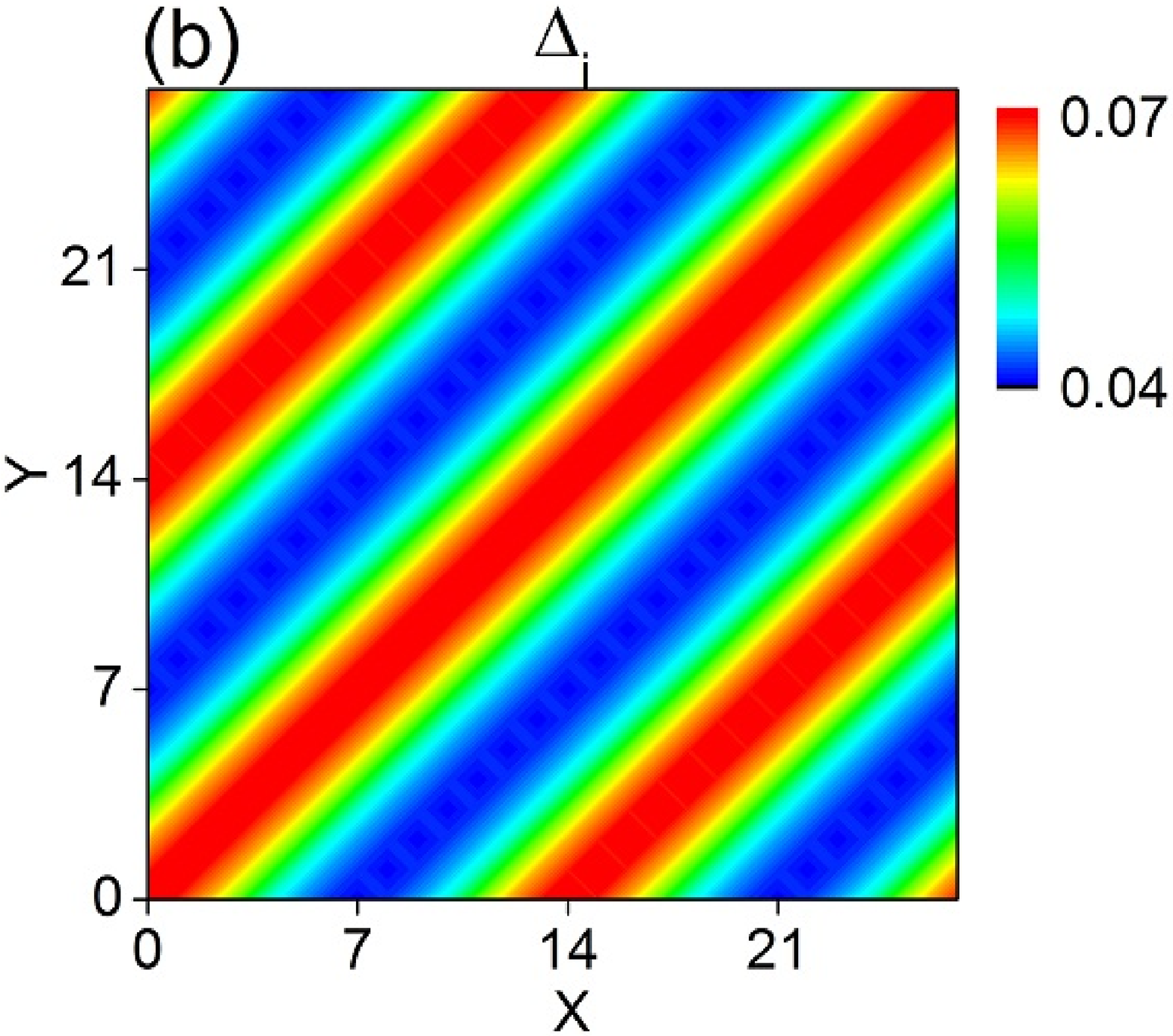}
\includegraphics[width=2.0in] {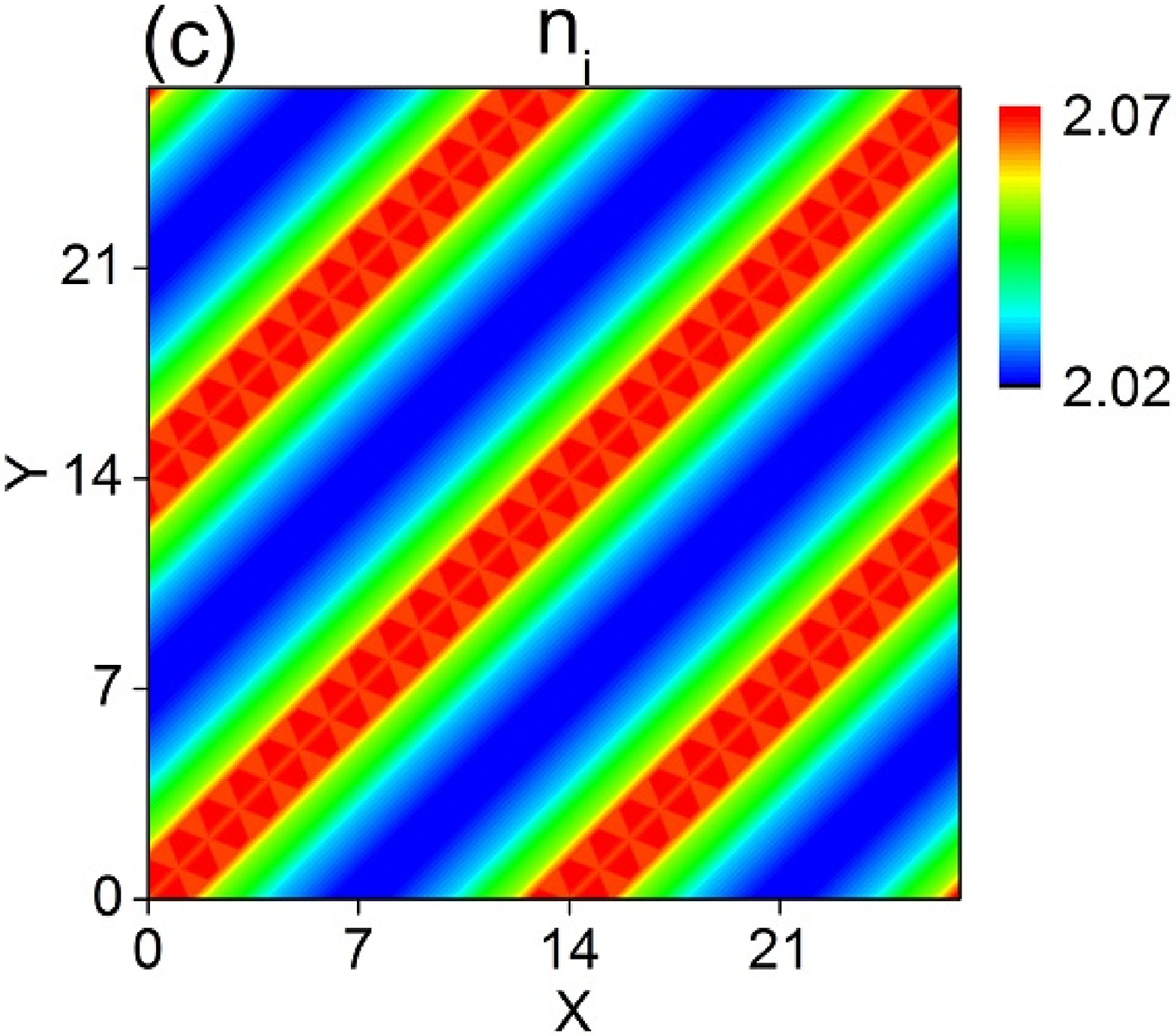}
\caption{Spatial profiles of ($a$) the magnetic order, ($b$) the
superconducting order, and ($c$) the charge density order are
presented for type-1 TB oriented $45^\circ$ from the
x-axis.}\label{Figa}
\end{figure}

TBs can  be formed by exchanging the lattice constants $a$ and $b$
on the opposite side of the TB. Figure \ref{Figa0}($a$) shows the
structure of a single such TB oriented at $45^\circ$ with respect to
the $x$-axis. Since the magnitude of $a$ is lightly larger that that
of $b$, this TB can be realized by assuming slightly different
nearest neighbor hoping terms $t_a=1.0$ and $t_b=1.2$ across the TB.
To analyze the effect of this TB, we considered a $28\times28$
lattice divided into 4 different domains separated by three parallel
TBs (see figure \ref{Figa0}($a$)) along the lines $y=x+14$, $y=x$
and $y=x-14$, in order to satisfy the periodic boundary conditions.
As shown in figure \ref{Figa0}($b$) and figure \ref{Figa}($a$),
there are three $90^\circ$ DWs formed and pinned on the TBs. The
patterns of these quantities are very similar to those found in
\cite{Huang} without the TBs (see figures 2($a$)-($c$) in
\cite{Huang}). However, in this case, the DW forms with a smaller
value of the Coulomb interaction $U$, indicating that the existence
of this type of TB is beneficial to the formation of the $90^\circ$
DWs. The solutions presented in figure \ref{Figa} are always stable
against the uniform $2\times1$ collinear AF order \cite{Zhou}.

Similar to the results without TBs \cite{Huang}, the SC as well as
the charge density get significantly enhanced on the DWs, which
occur at the TBs, and suppressed in the middle of the magnetic
domains (see figures \ref{Figa}($b$) and ($c$)). This conclusion is
consistent with the enhancement of SC along the TB as observed by
the SQIDM experiments \cite{Kalisky} on underdoped samples. There
exists another experiment \cite{Kalisky1} favoring our  result in
which the magnetic vortices are found not to be pinned at the TBs.
This implies that the SC is also enhanced along the TBs. For the
parent compound Ca(Fe$_{1-x}$Co$_{x}$As)$_{2}$ with $x\sim0$, our
numerical results show that the magnetic structures as shown in
figure \ref{Figa}($a$) still remains, but there is no SC across the
sample including the DWs, and this is in agreement with the STM
experiments \cite{Chuang}.

It is important to note that the lattices on both sides of this TB
should be well matched at the TB, and each of the unit cells along
the TB is only slightly deformed from the square shape. Thus, we do
not expect that scattering of the electrons from the lattice
mismatch across the TB should be strong.

\subsection{B. Type-1 Twin-Boundary Oriented $90^\circ$ From The x-Axis}
\begin{figure}[t]
\centering
\includegraphics[width=2.5in] {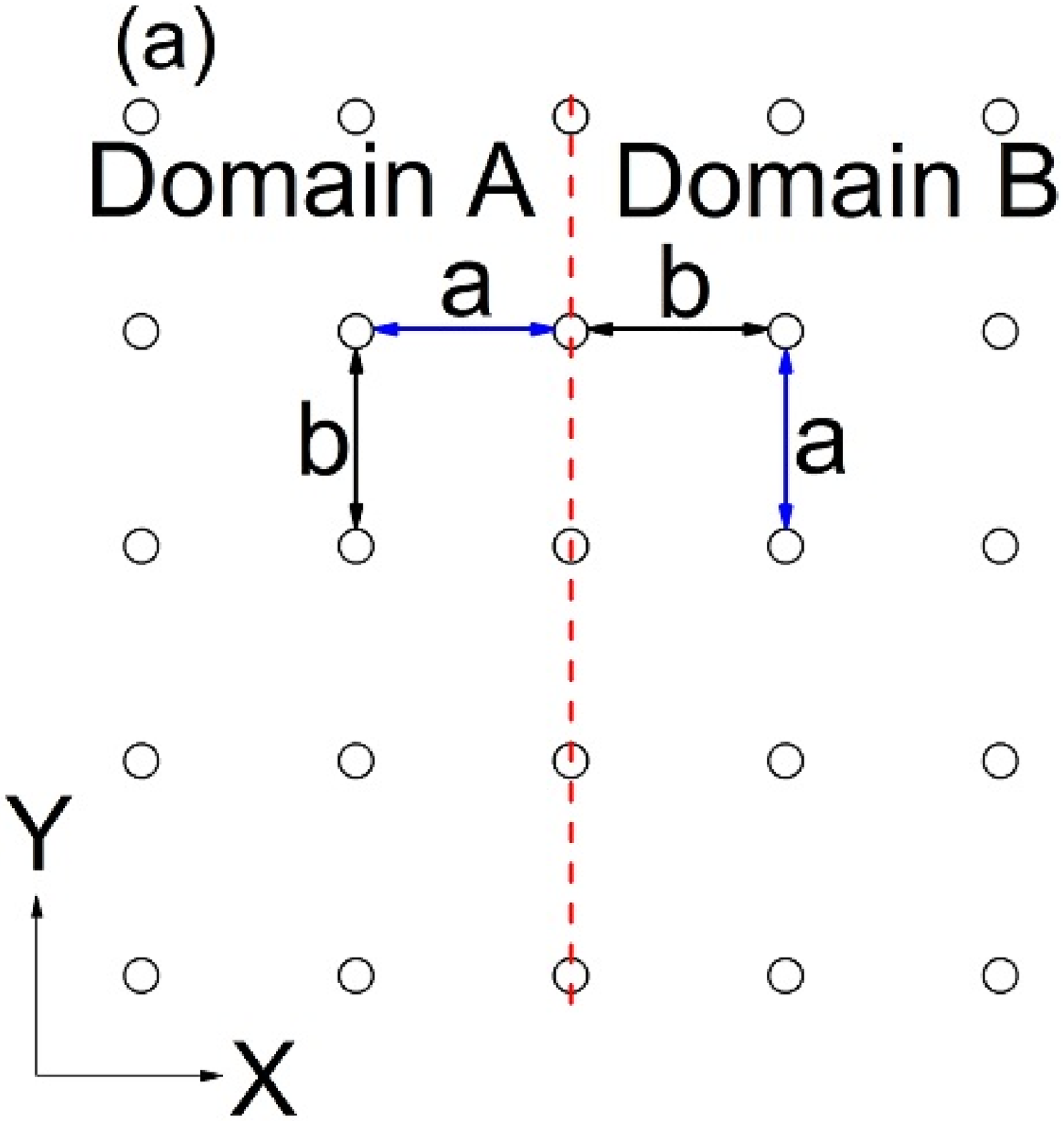}
\includegraphics[width=2.5in] {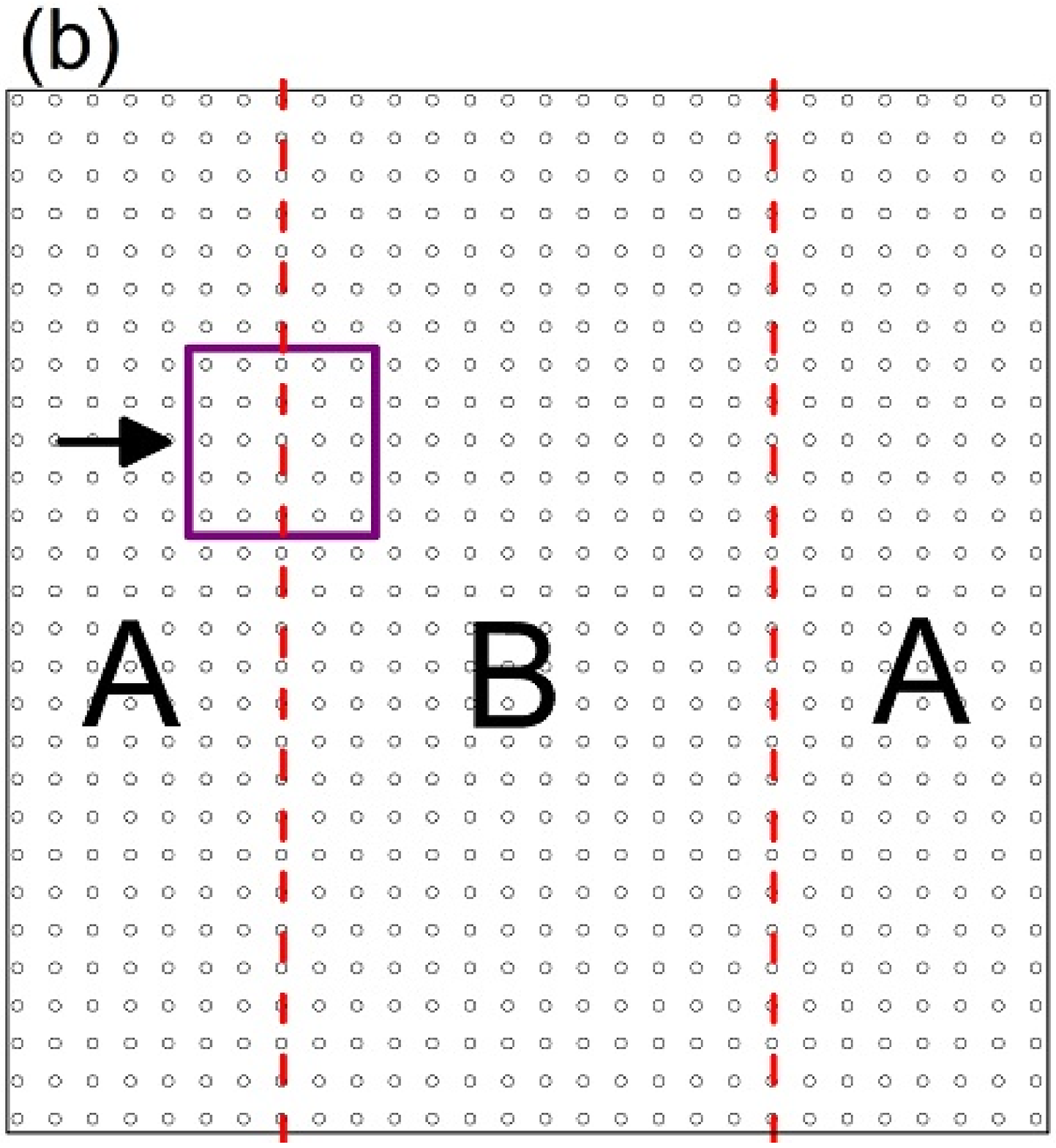}
\caption{($a$) The lattice structure near a TB parallel to the
$y$-direction (red dashed line), the open circles represent the
positions of Fe atoms, a (blue solid line) and b (black solid line)
are the lattice constants along x and y directions in domain A.
($b$) Two TBs in the $28\times28$ lattice, the black arrow indicates
the position of the lattice structure shown in ($a$) in the whole
lattice.} \label{Figb0}
\end{figure}

A TB formed by exchanging the $a$ and $b$ lattice constants across
it can also be oriented parallel to the $x$- or $y$-axis (as shown
in figure \ref{Figb0}($a$)). The periodic boundary condition of this
case is achieved by dividing the system into 3 different domains
separated by two TBs (see figure \ref{Figb0}($a$)) along the lines
$x=7$ and $x=20$, as shown in figure \ref{Figb0}($b$).

\begin{figure}[t]
\centering
\includegraphics[width=2.0in] {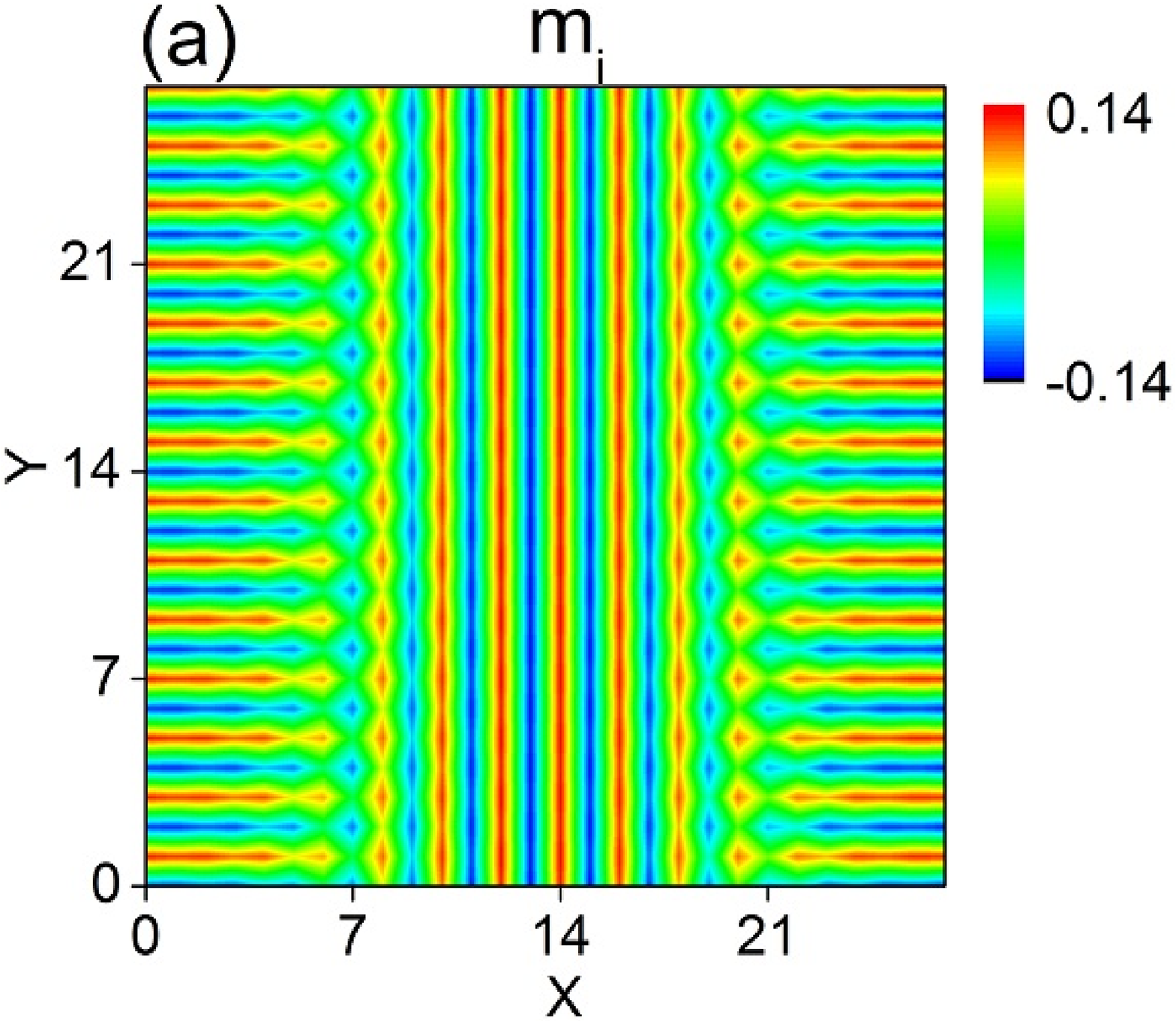}
\includegraphics[width=2.0in] {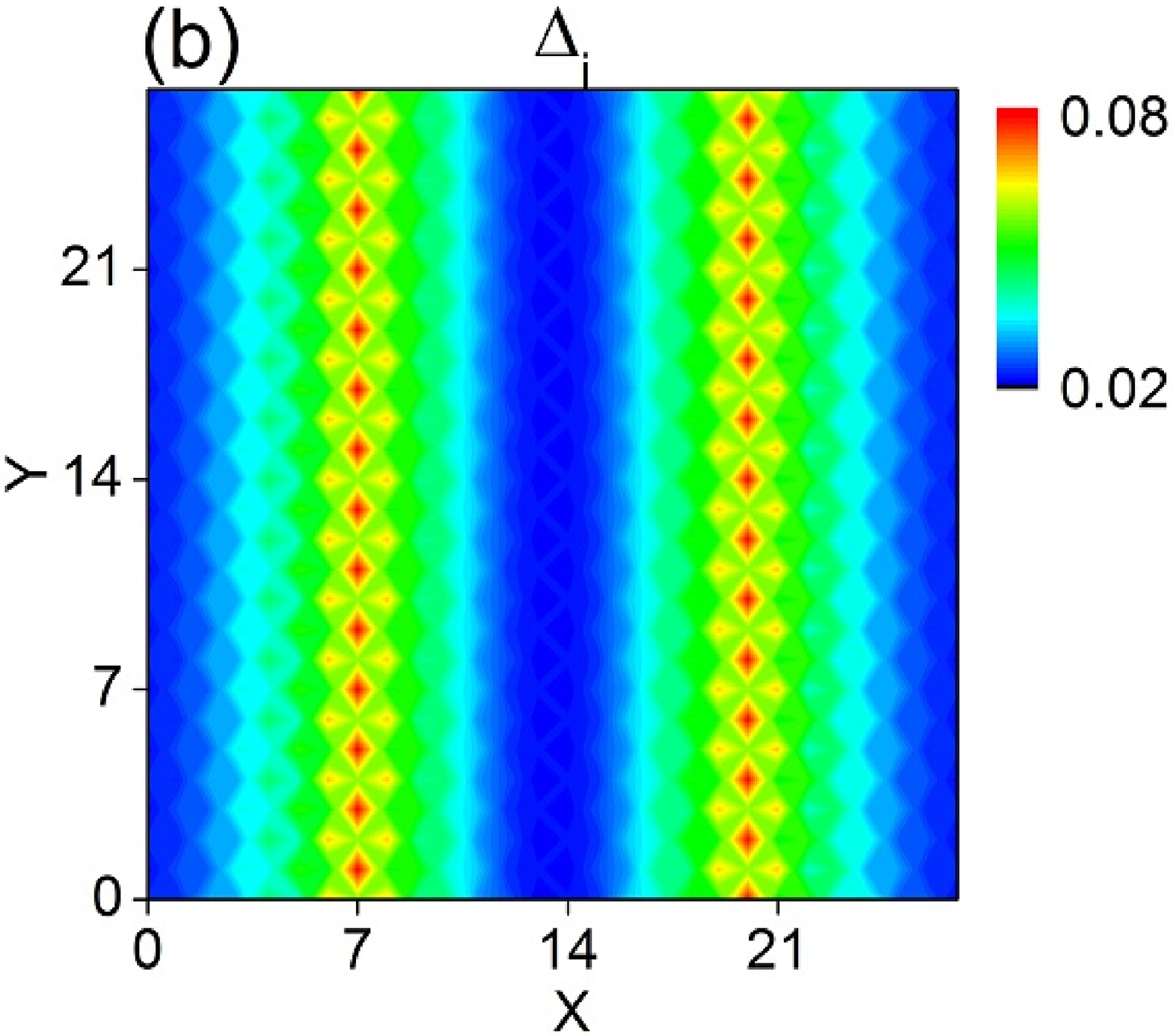}
\includegraphics[width=2.0in] {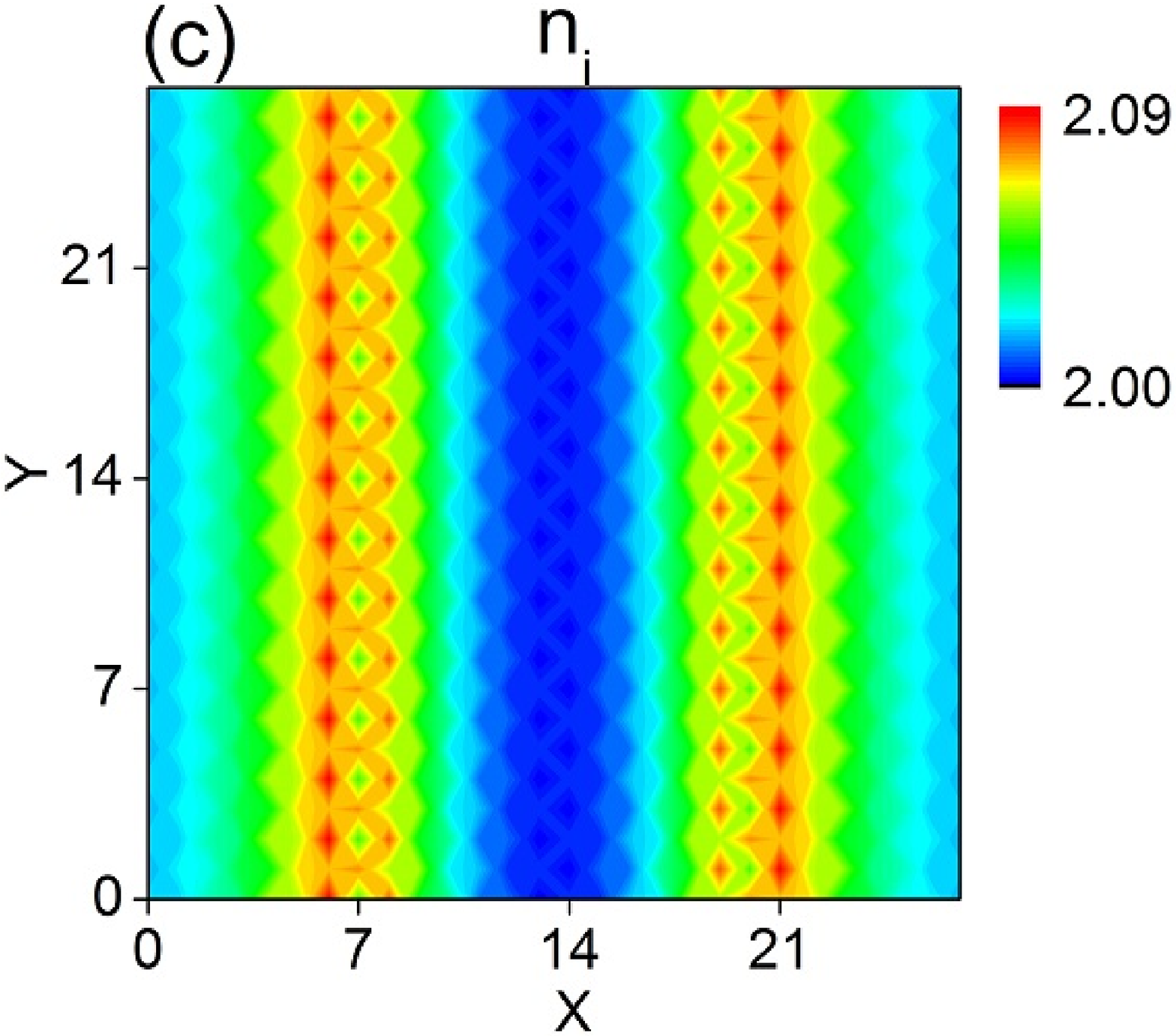}
\caption{Spatial profiles of ($a$) the magnetic order, ($b$) the
superconducting order, and ($c$) the charge density order are
presented for type-1 TB oriented $90^\circ$ from the x-axis.}
\label{Figb}
\end{figure}

Here the magnetic DWs are pinned at the TBs (see figure
\ref{Figb}($a$)) on which existing a weak local ferromagnetic order
which a periodic modulation. The SC also has a similar periodic
modulation and is enhanced on the DWs, but suppressed in the middle
of the magnetic domains (see figure \ref{Figb}($b$)). A charge
density wave appears near the DWs (see figure \ref{Figb}($c$)) while
the electron density gets suppressed in the middle of the magnetic
domains. It is important to point out that in this case, the
lattices on the opposite sides of a TB are not well matched.
Therefore, there may be considerable scattering of the electrons due
to the lattice mismatches near these TBs. If this effect is
included, we expect that the SC would get suppressed, instead of
being enhanced when the mismatch becomes strong across the DWs or
the TBs. This issue will be discussed in Section 5.

\subsection{C. Type-2 Twin-Boundary Oriented $90^\circ$ From The x-Axis}
\begin{figure}[t]
\centering
\includegraphics[width=2.5in] {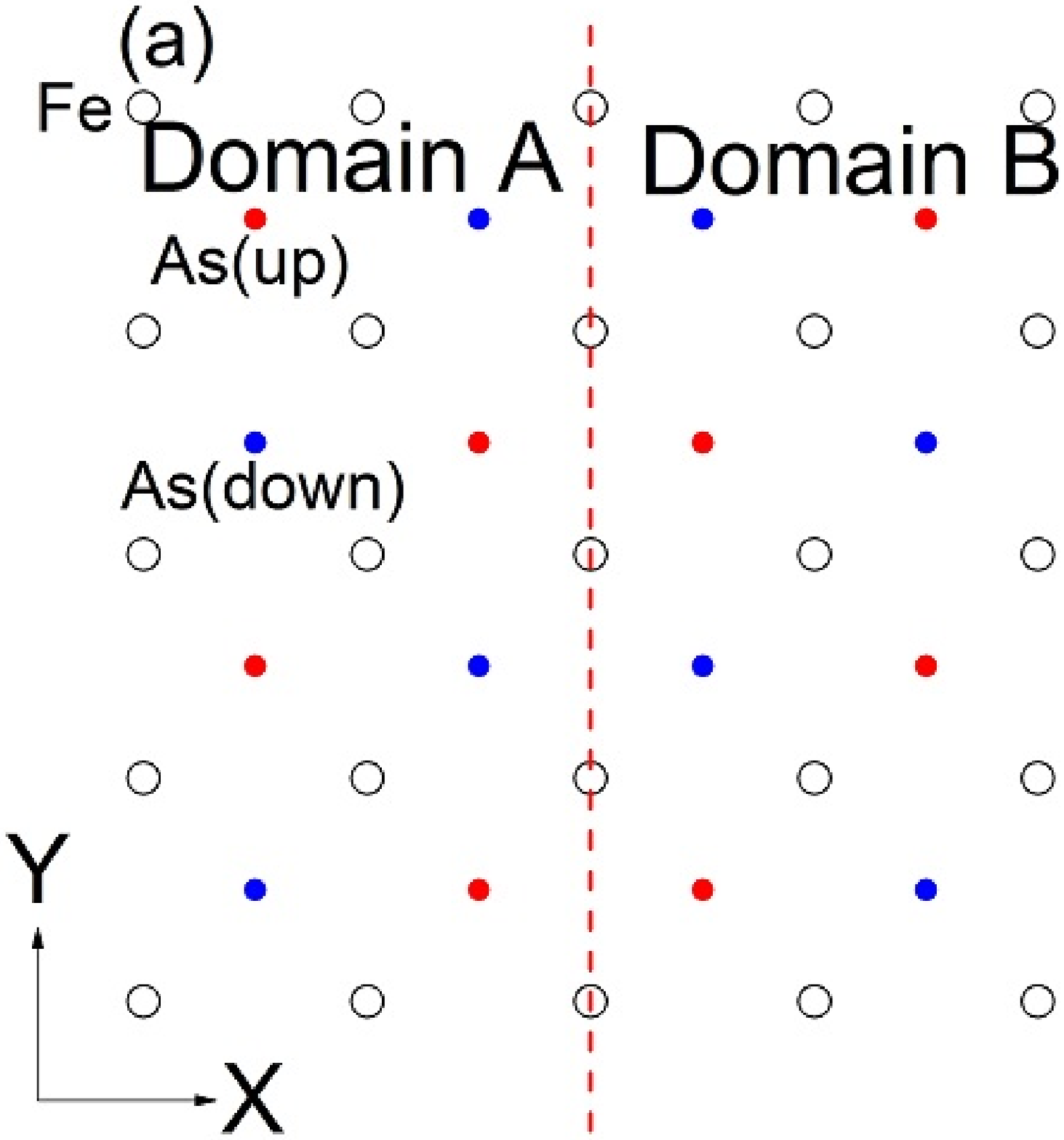}
\includegraphics[width=2.5in] {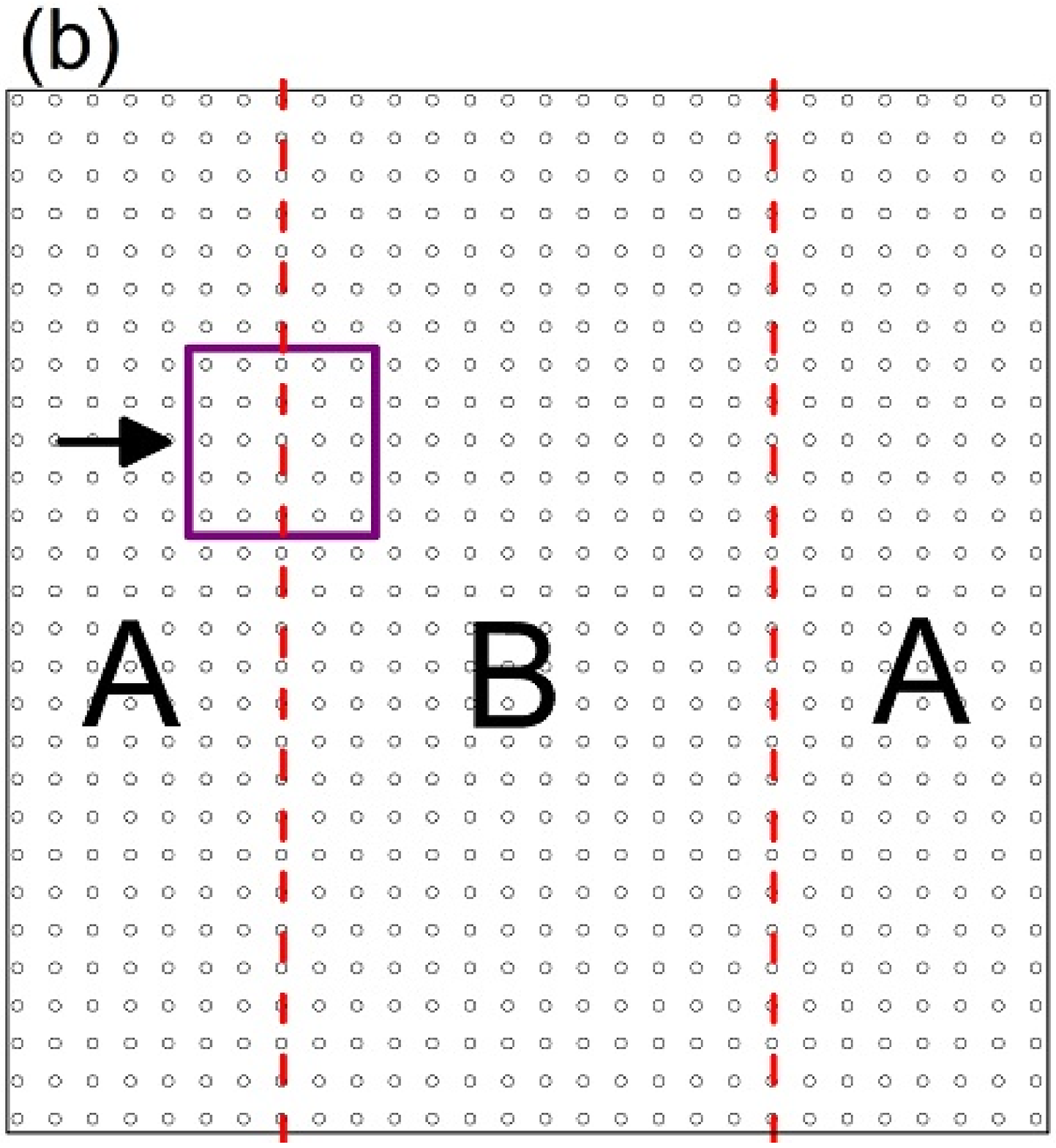}
\caption{($a$) The lattice structure near a twin boundary (red
dashed line) parallel to the $y$-direction formed by misplacing As
atoms. The open circles represent the positions of Fe atoms, and the
red and blue dots respectively denote the As(up) and As(down) atoms.
($b$) Two TBs in the $28\times28$ lattice, the black arrow indicates
the position of the lattice structure shown in ($a$) in the whole
lattice.} \label{Figc0}
\end{figure}

\begin{figure}[t]
\centering
\includegraphics[width=2.0in] {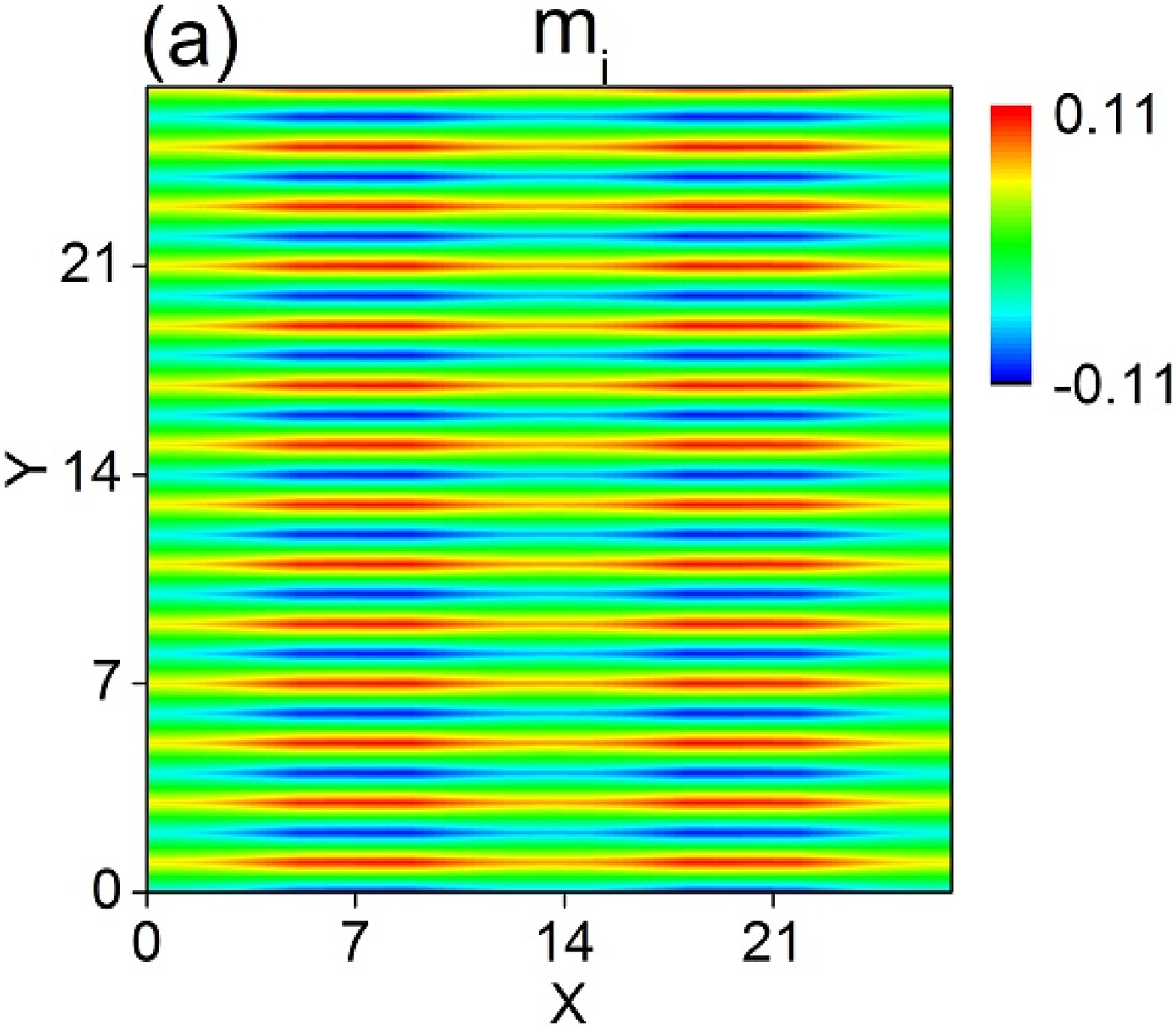}
\includegraphics[width=2.0in] {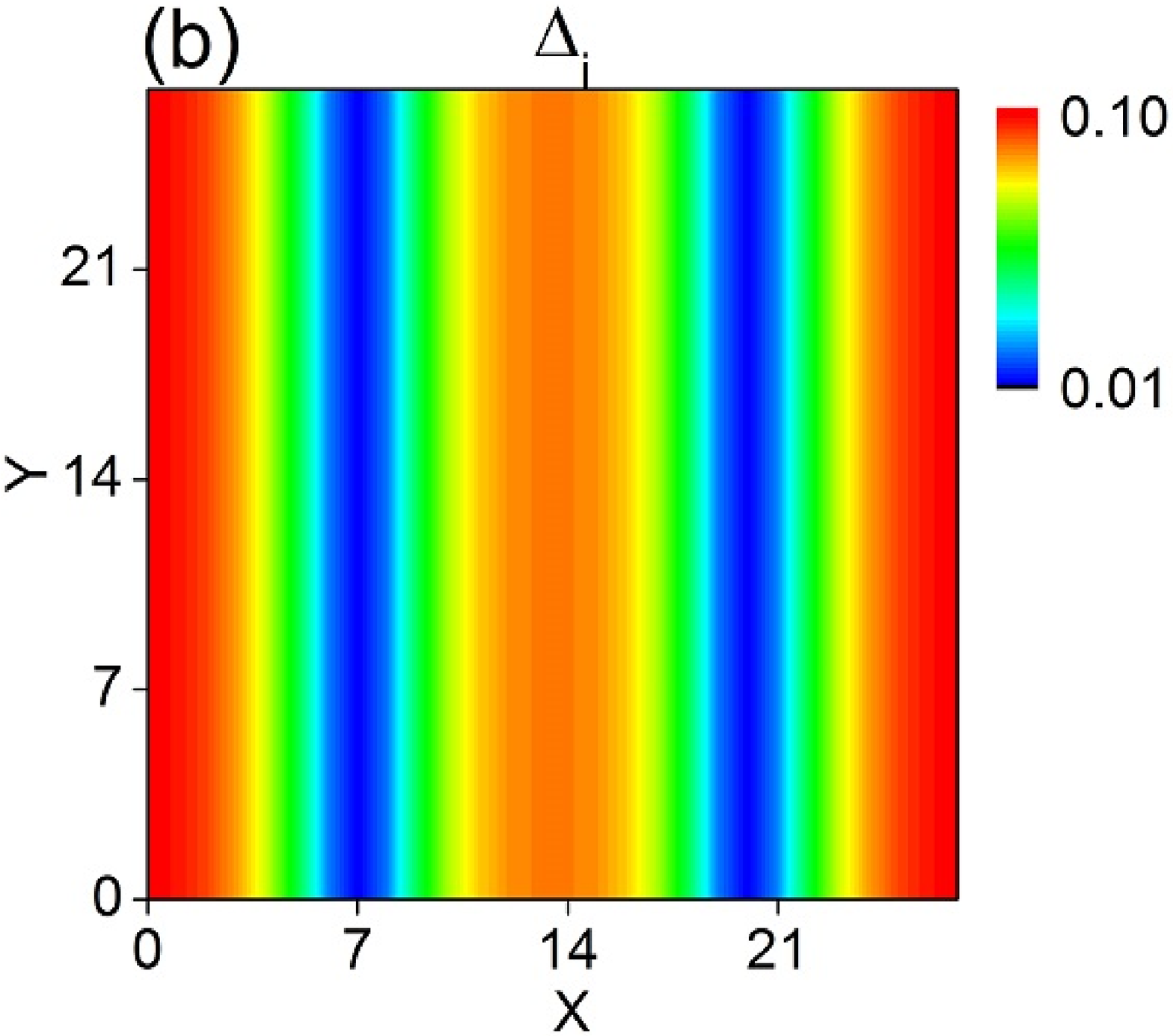}
\includegraphics[width=2.0in] {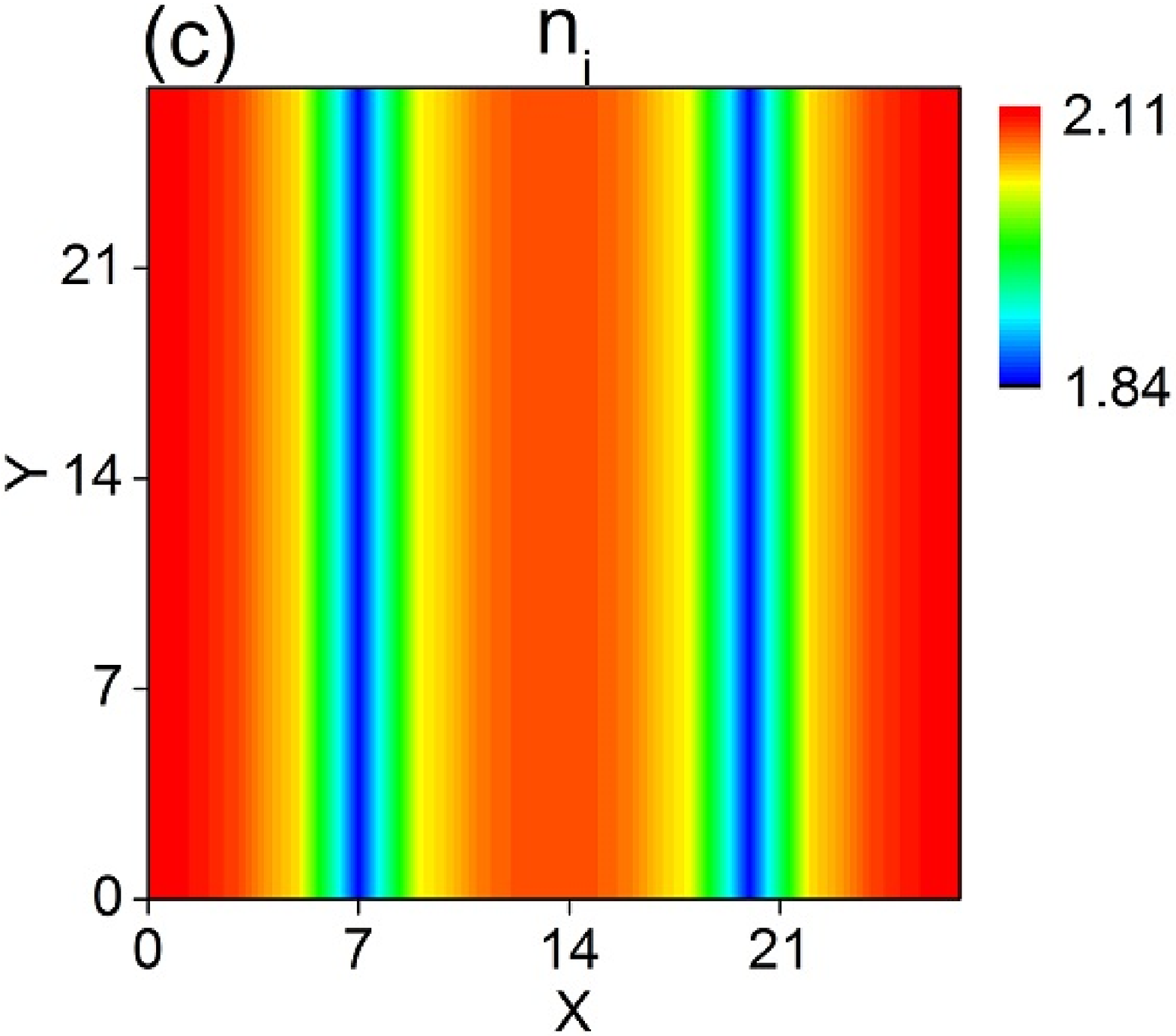}
\caption{Spatial profiles of ($a$) the magnetic order, ($b$) the
superconducting order, and ($c$) the charge density order are
presented for type-2 TB oriented $90^\circ$ from the x-axis.}
\label{Figc}
\end{figure}

Another possible TB can be generated by slipping the lattice on the
right side of the TB by a lattice constant along the y-direction
with respect to the lattice on the left of the TB. There are two
different types of As atoms in our model, we label them as As(up)
and As(down) atoms which are asymmetrically place above and below
the Fe plane. This type of TBs can be clearly seen from figure
\ref{Figc0}($a$), in which the TB is represented by the red-dashed
line. The crystal lattice for the FeAs layer has D$_{2d}$ symmetry,
namely the 4 nearest neighboring As atoms of a "down" As atom should
be all "up". The hopping terms between the next-nearest-neighboring
Fe ions via the hybridization of the 4p orbital with the As atom in
the middle should have different values depending on whether the As
atom is above ($t_2$) or below ($t_3$) the Fe plane \cite{Zhang,Hu}.
The D$_{2d}$ symmetry is broken by the presence of the TB. We
considered a $28\times28$ lattice with periodic boundary conditions
divided into three domains by two TBs located at $x=7$ and $x=20$
(see figure \ref{Figc0}($b$)). Figure \ref{Figc}($a$) shows the
magnetic order is enhanced near the TBs. Defining the magnetic DWs
to be where the magnetic order is suppressed, that is, near the
middle between two TBs. Then, clearly, the DWs are not located at
the TBs. On the opposite sides of a DW, there is no change in the
magnetic phase, thus we could label the DWs as the 0-phase DWs.
Figure \ref{Figc}($b$) shows that the SC is enhanced along the DWs,
and that it is suppressed near the TBs. Figure \ref{Figc}($c$) shows
that the electron density is depleted near the TBs and becomes
lightly hole doped. Apparently, the depleted electron density leads
to strong magnetic order that suppresses the SC order. On the DWs,
the electron density appears to be close to optimal doping and thus
SC gets enhanced and the magnetic order is suppressed. We shall also
point out in Section 6 that the magnetic and SC structures near a TB
oriented $0^\circ$ from the x-axis are very different from the
present case, and anti-phase DWs are predicted.

\subsection{D. Type-2 Twin-Boundary Oriented $45^\circ$ From The x-Axis}
\begin{figure}[t]
\centering
\includegraphics[width=2.5in] {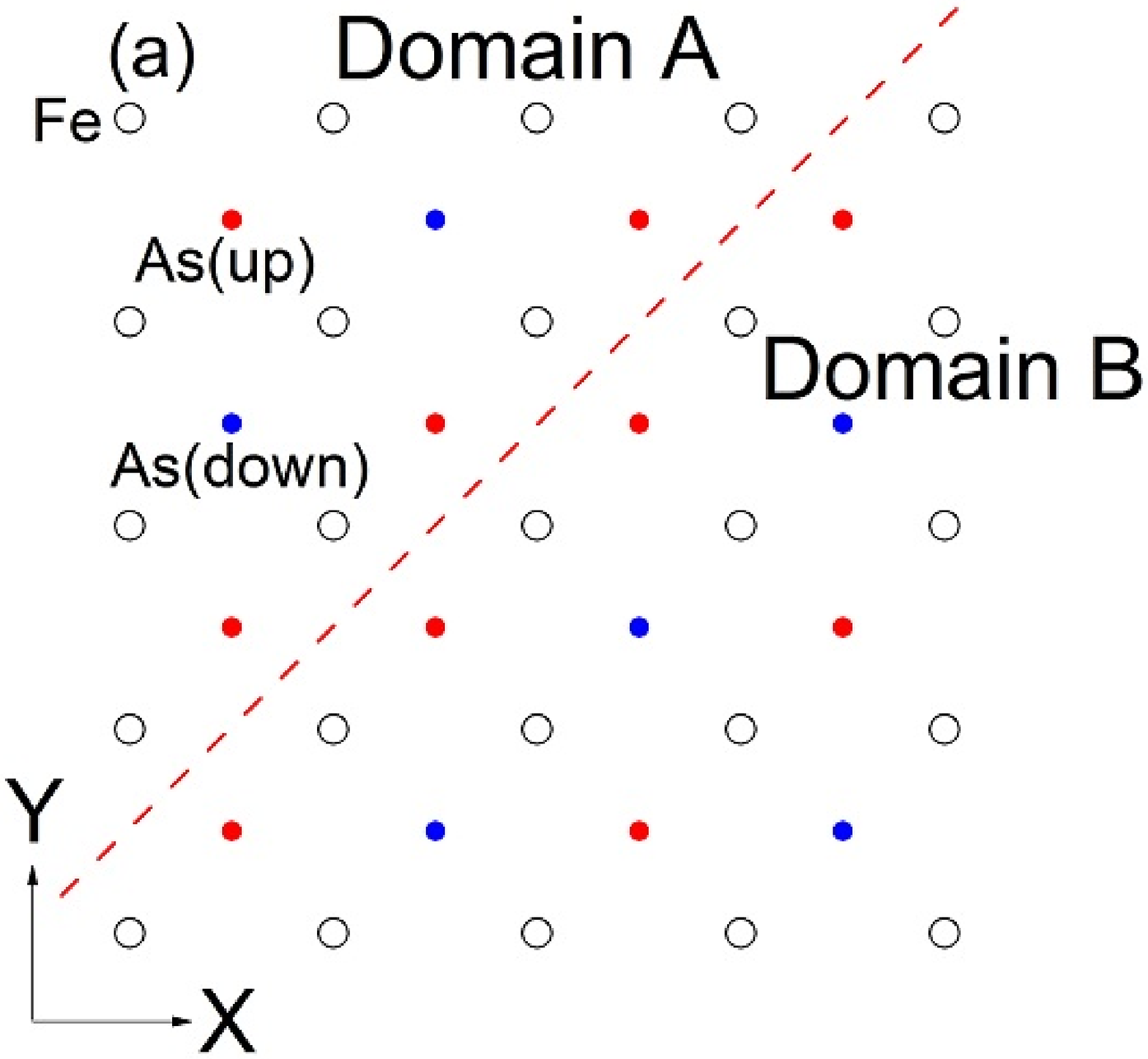}
\includegraphics[width=2.5in] {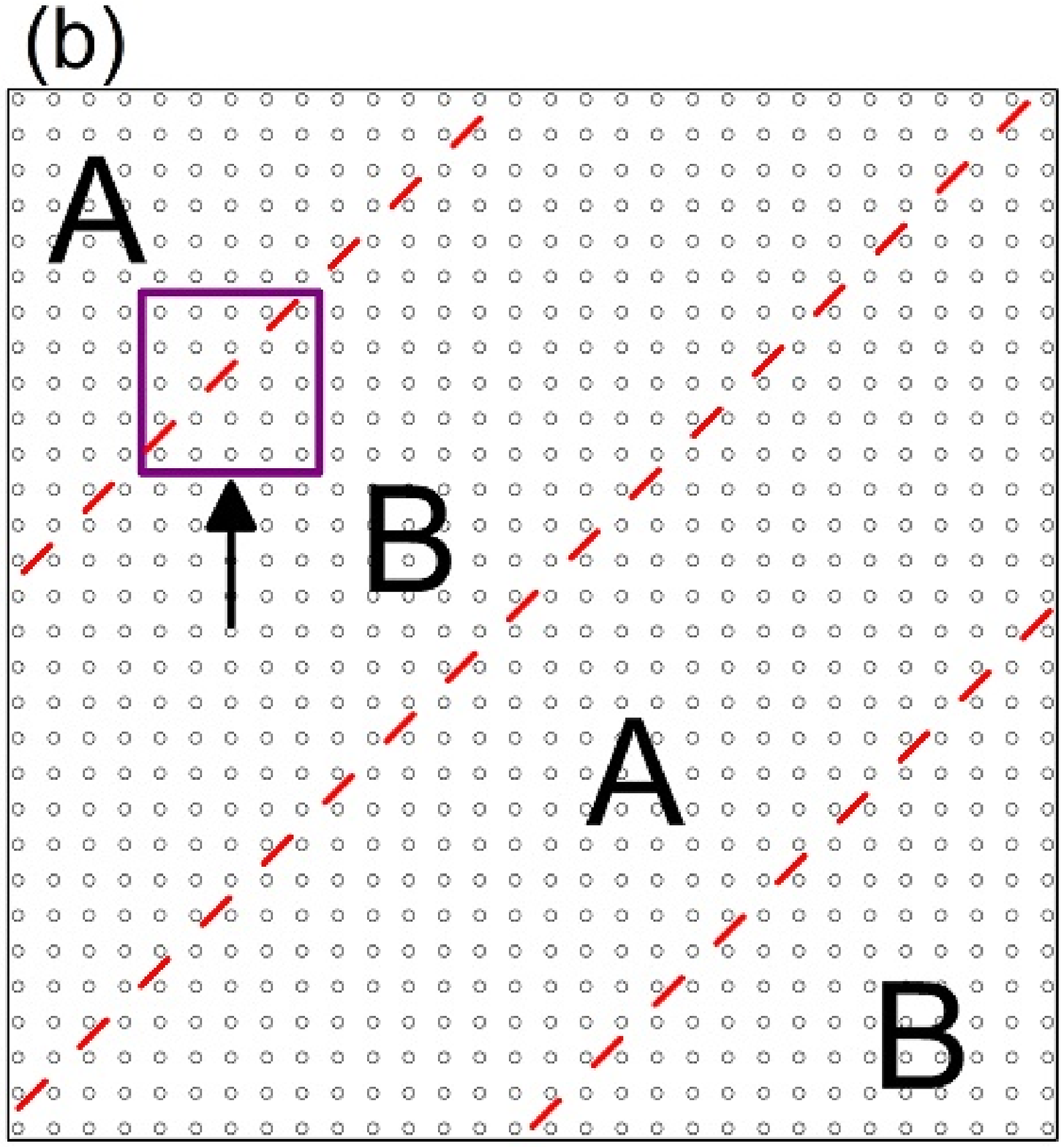}
\caption{($a$) The lattice structure near a twin boundary (red
dashed line) with misplacing the As atoms along diagonal (or
$45^\circ$) direction. The open circles represent the positions of
Fe atoms, and the red and blue dots respectively denote the As(up)
and As(down) atoms. ($b$) Three TBs in the $30\times30$ lattice, the
black arrow indicates the position of the lattice structure shown in
($a$) in the whole lattice.} \label{Figd0}
\end{figure}

A TB due to missing one line of the lattice which contains both Fe
and As(down) atoms oriented along $45^\circ$ from the x-axis is
shown in figure \ref{Figd0}($a$). In fact this "TB" could also be
regarded a line of missing Fe-As atoms. The $D_{2d}$ symmetry of the
lattice is also broken by the presence of this TB. Note that the
geometry of this TB is fundamentally different from the one showed
in figure \ref{Figc0}($a$) since the TB does not pass through any of
the Fe or As atoms. To study this case, we considered a $30\times30$
lattice and three identical TBs oriented $45^\circ$ from the x-axis
(see figure \ref{Figd0}($b$)) in order to satisfy the periodic
boundary condition. The TBs are located along $y=x-15$, $y=x$ and
$y=x+15$.

\begin{figure}[t]
\centering
\includegraphics[width=2.0in] {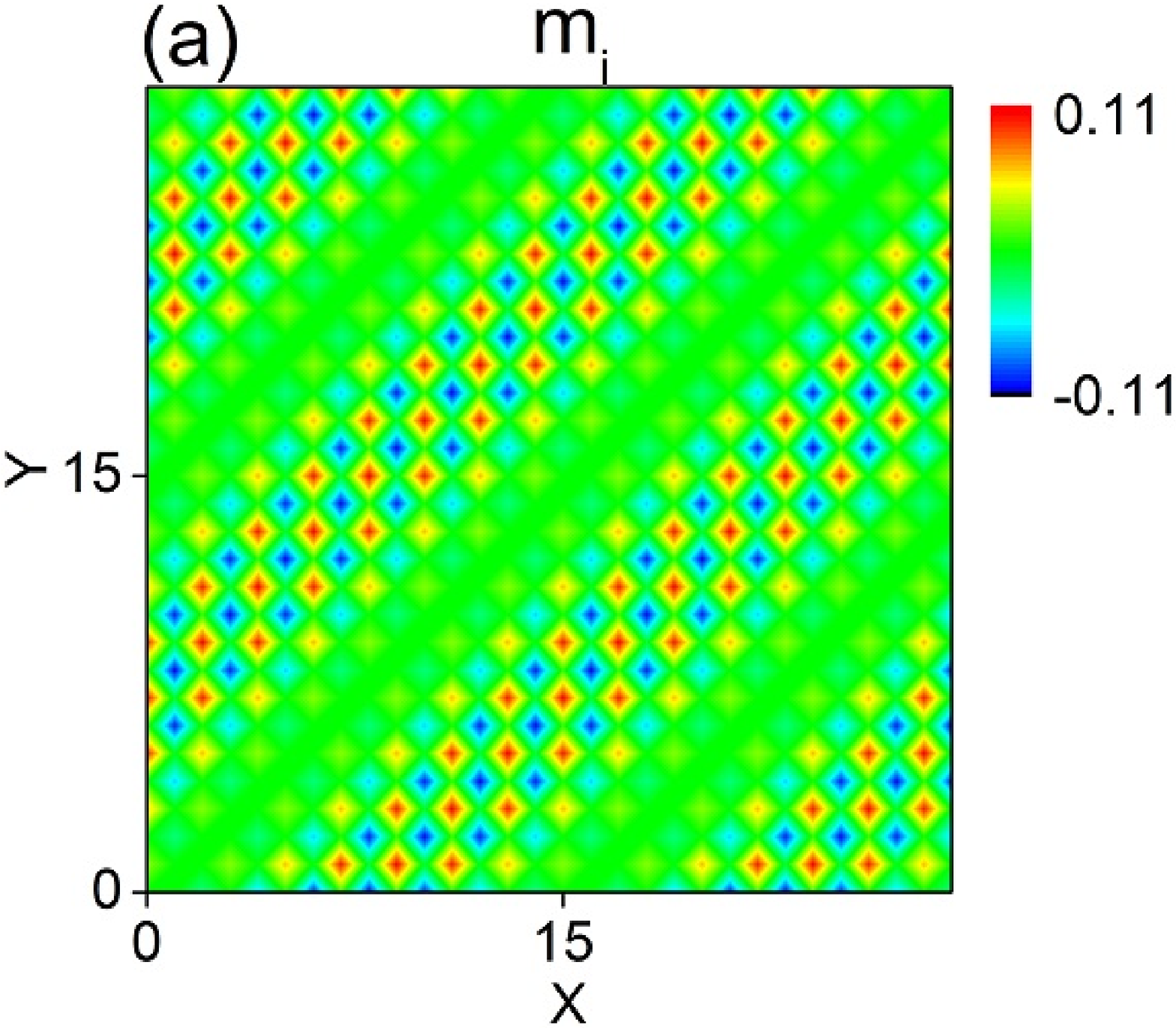}
\includegraphics[width=2.0in] {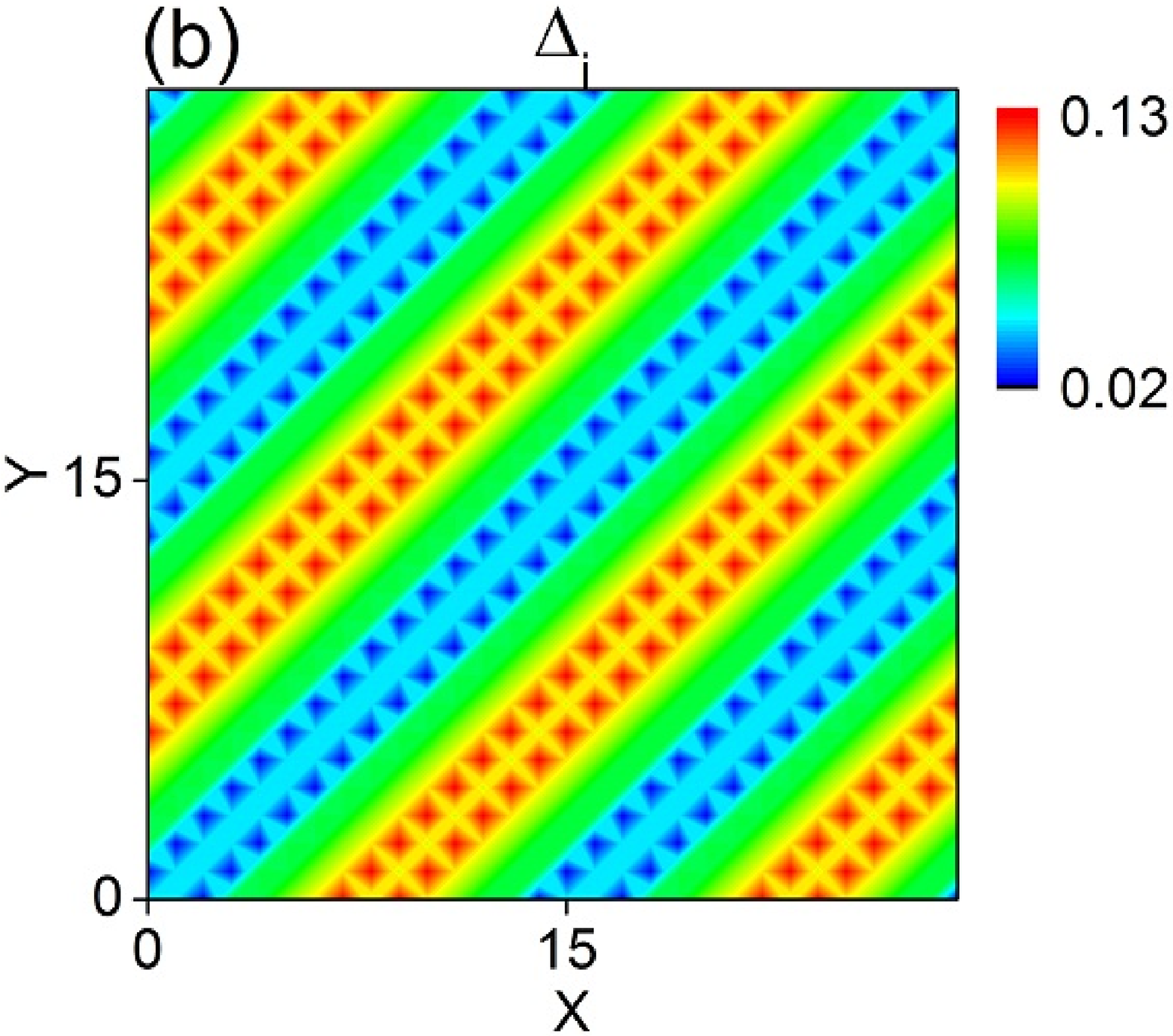}
\includegraphics[width=2.0in] {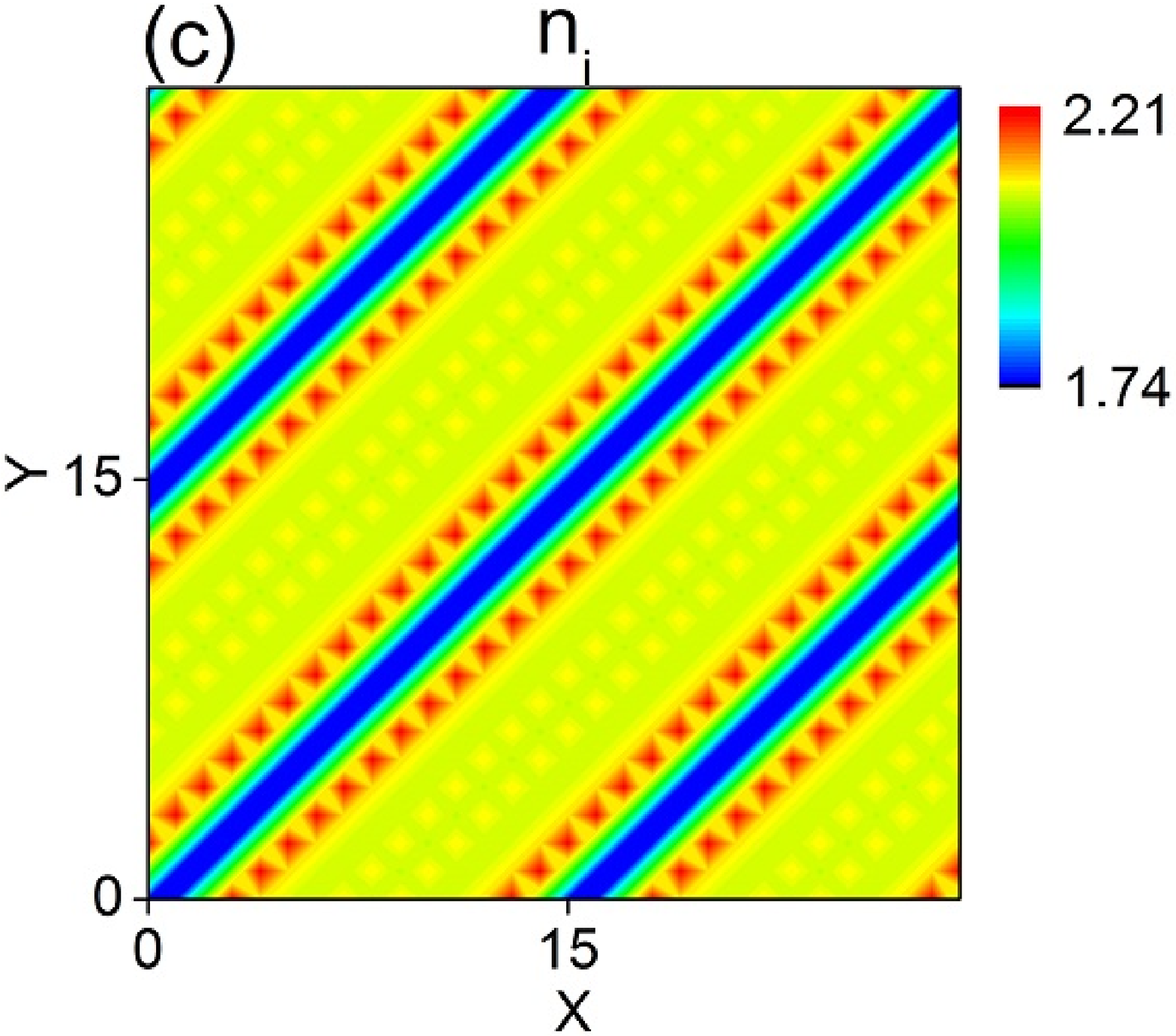}
\caption{Spatial profiles of ($a$) the magnetic order, ($b$) the
superconducting order, and ($c$) the charge density order are
presented for type-2 TB oriented $45^\circ$ from the x-axis.}
\label{Figd}
\end{figure}

Different from the case in figure \ref{Figc}($a$), the magnetic
order shown in figure \ref{Figd}($a$) is suppressed along the TBs
where DWs are located. The magnetic domain between the TBs still has
the usual $2\times1$ collinear AF structure, except the magnetic
moments are strongly and periodically modulated along the x-axis,
which may be due to finite size effects of the TB and the small
distance between two nearest neighboring TBs. It also appears that
the local $2\times1$ collinear AF structure could be represented by
a stripe-like $\sqrt{2}\times\sqrt{2}$ AF structure oriented
$45^\circ$ from the x-axis. Furthermore, the SC is enhanced in these
regions (see figure \ref{Figd}($b$)). From figure \ref{Figd}($c$),
we can find that on the TBs or DWs the carrier density is
corresponding to that in the overly hole-doped case
($x\approx-0.3$), which explains why the magnetic and SC orders are
suppressed on both sides of the TB. Interestingly, stripe-like
charge density waves oriented $45^\circ$ from the x-axis occur on
both sides of each TB.

\begin{figure}[t]
\centering
\includegraphics[width=2.0in] {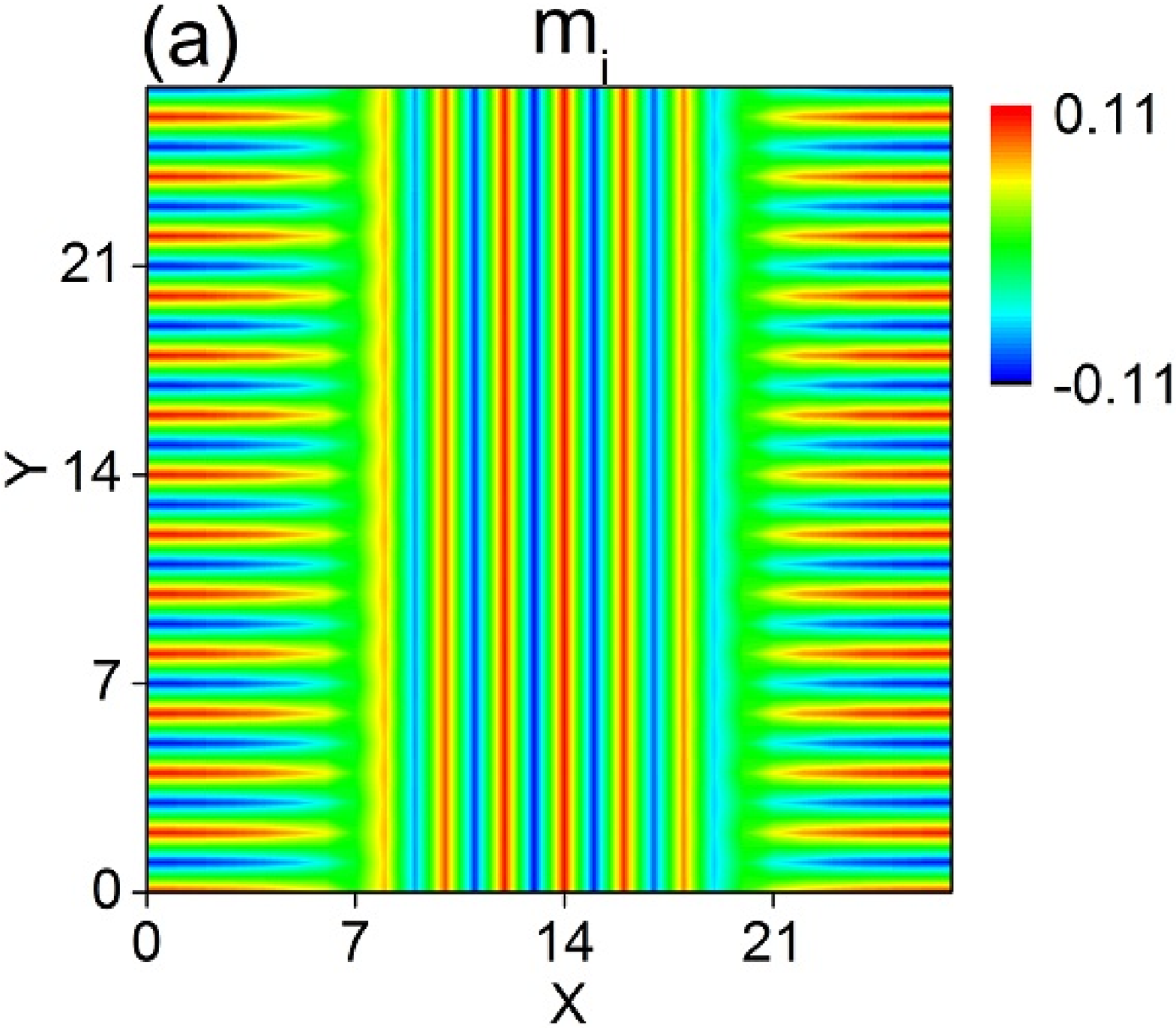}
\includegraphics[width=2.0in] {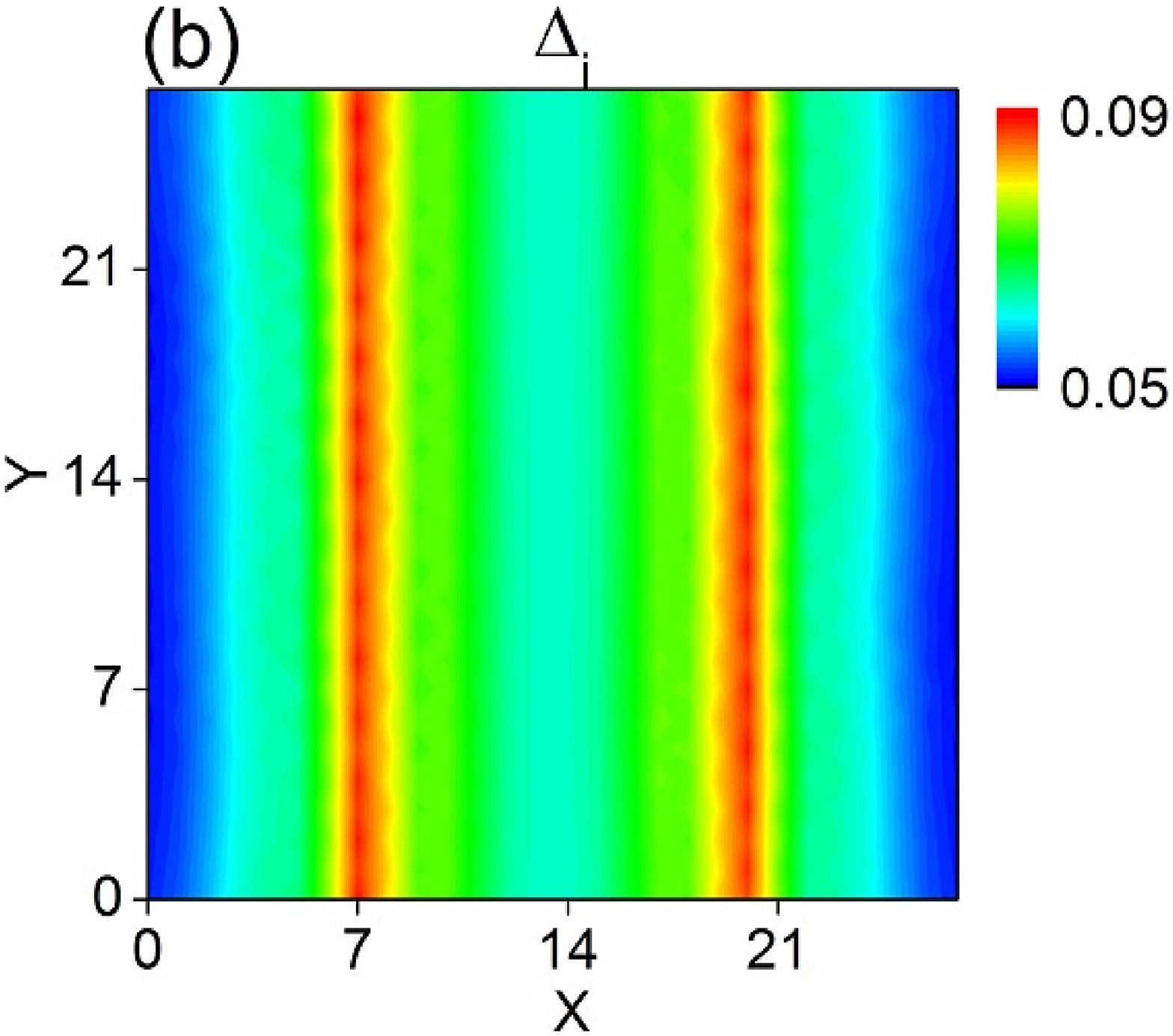}
\includegraphics[width=2.0in] {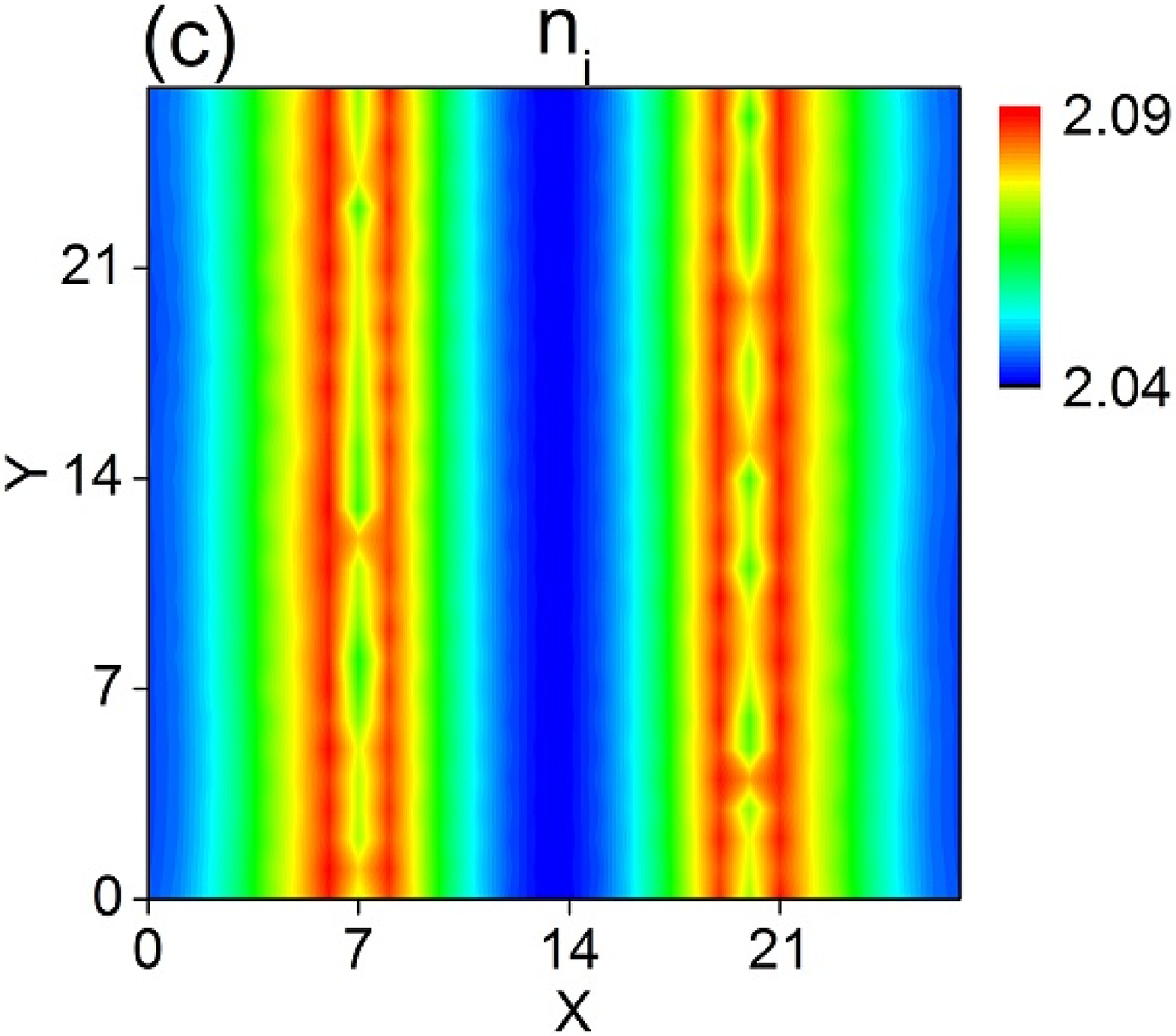}
\includegraphics[width=2.0in] {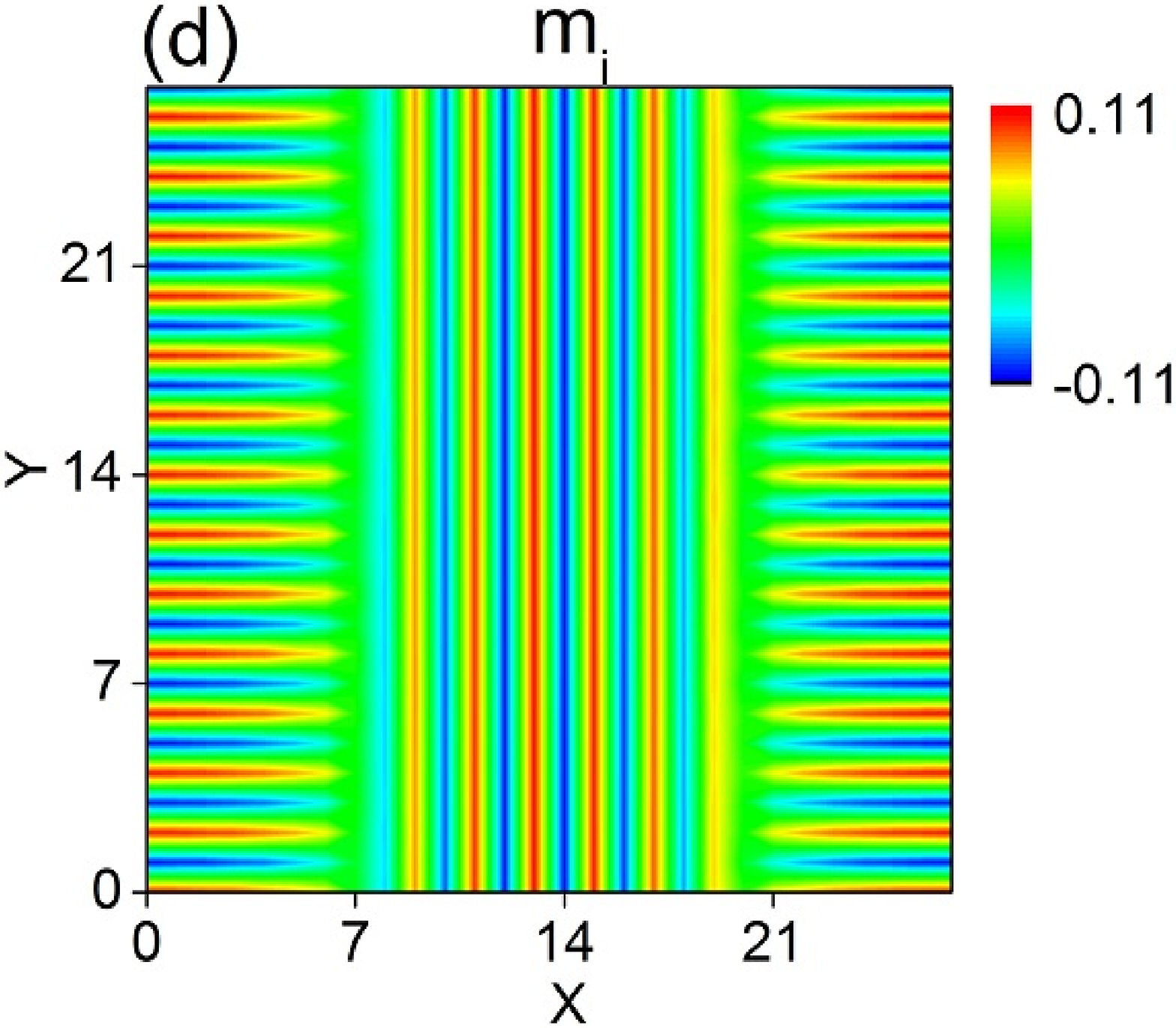}
\includegraphics[width=2.0in] {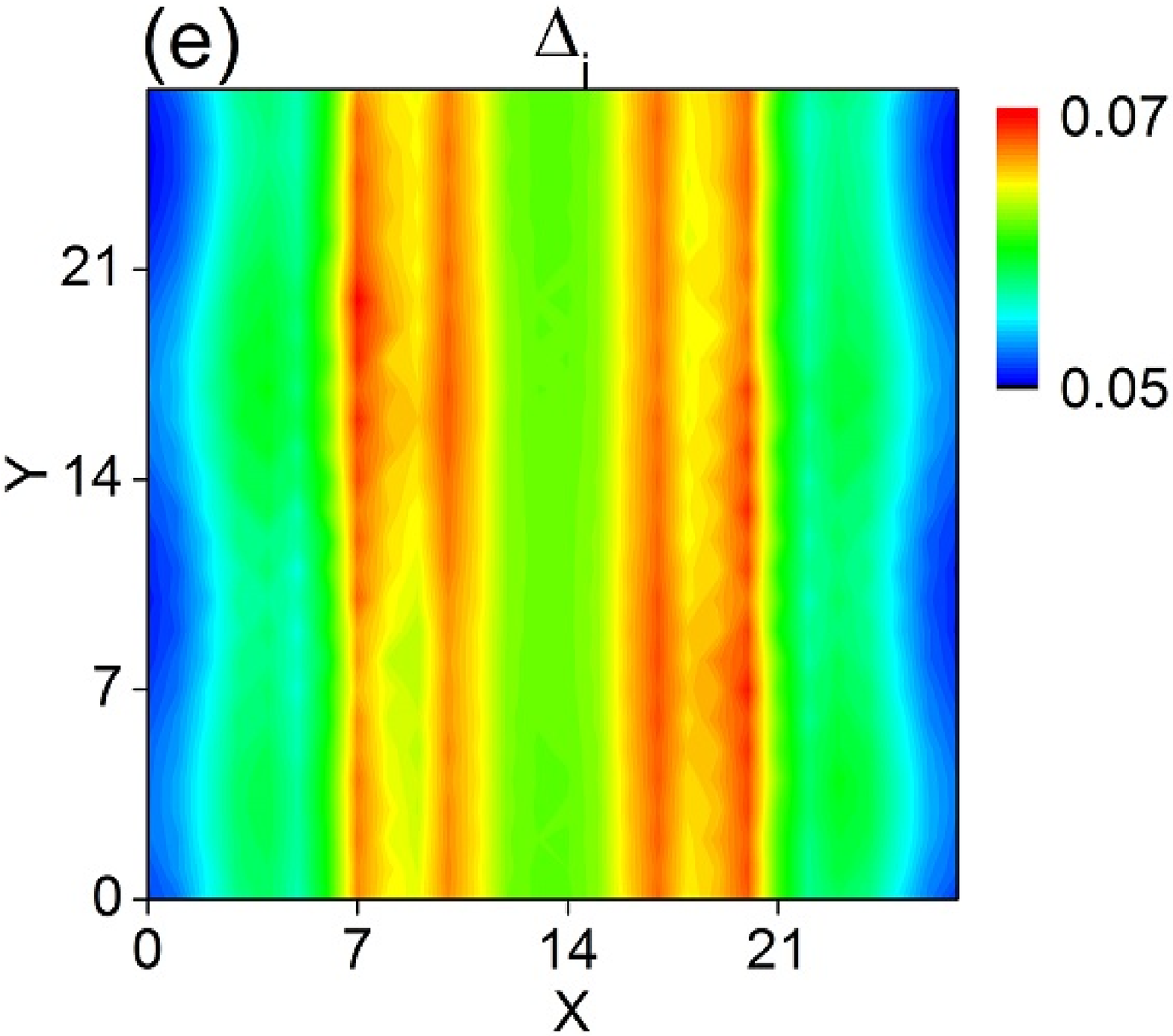}
\includegraphics[width=2.0in] {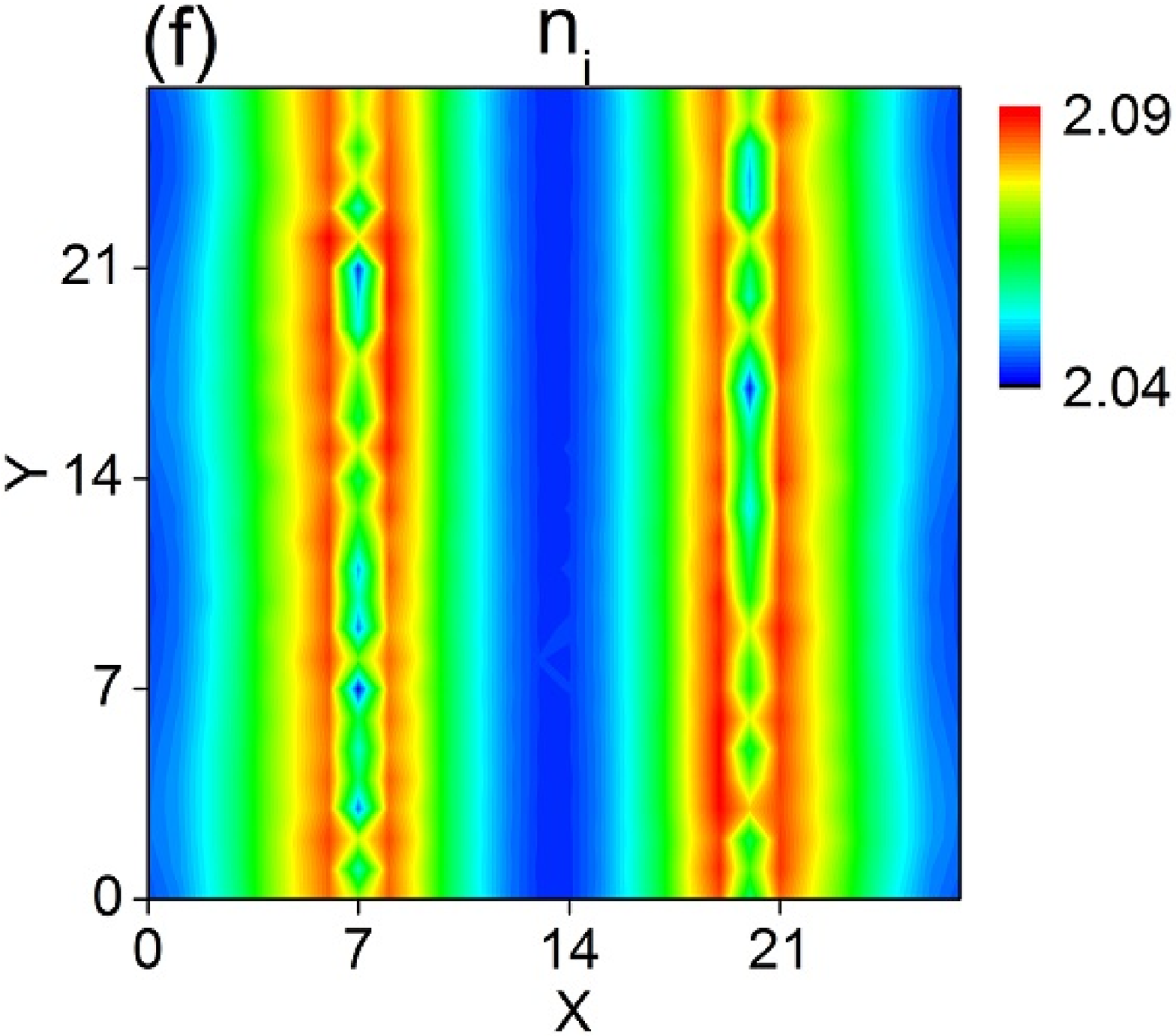}
\includegraphics[width=2.0in] {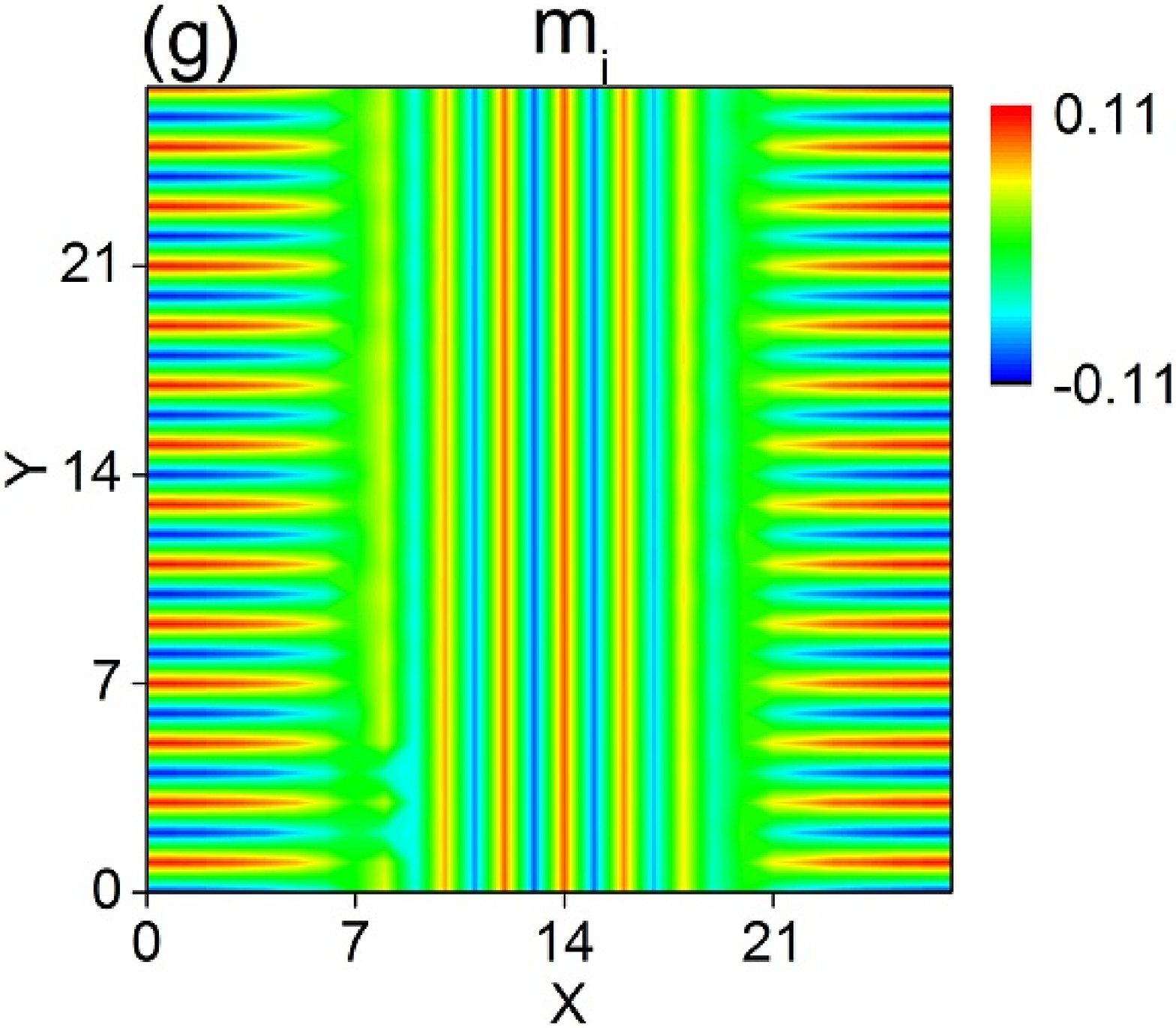}
\includegraphics[width=2.0in] {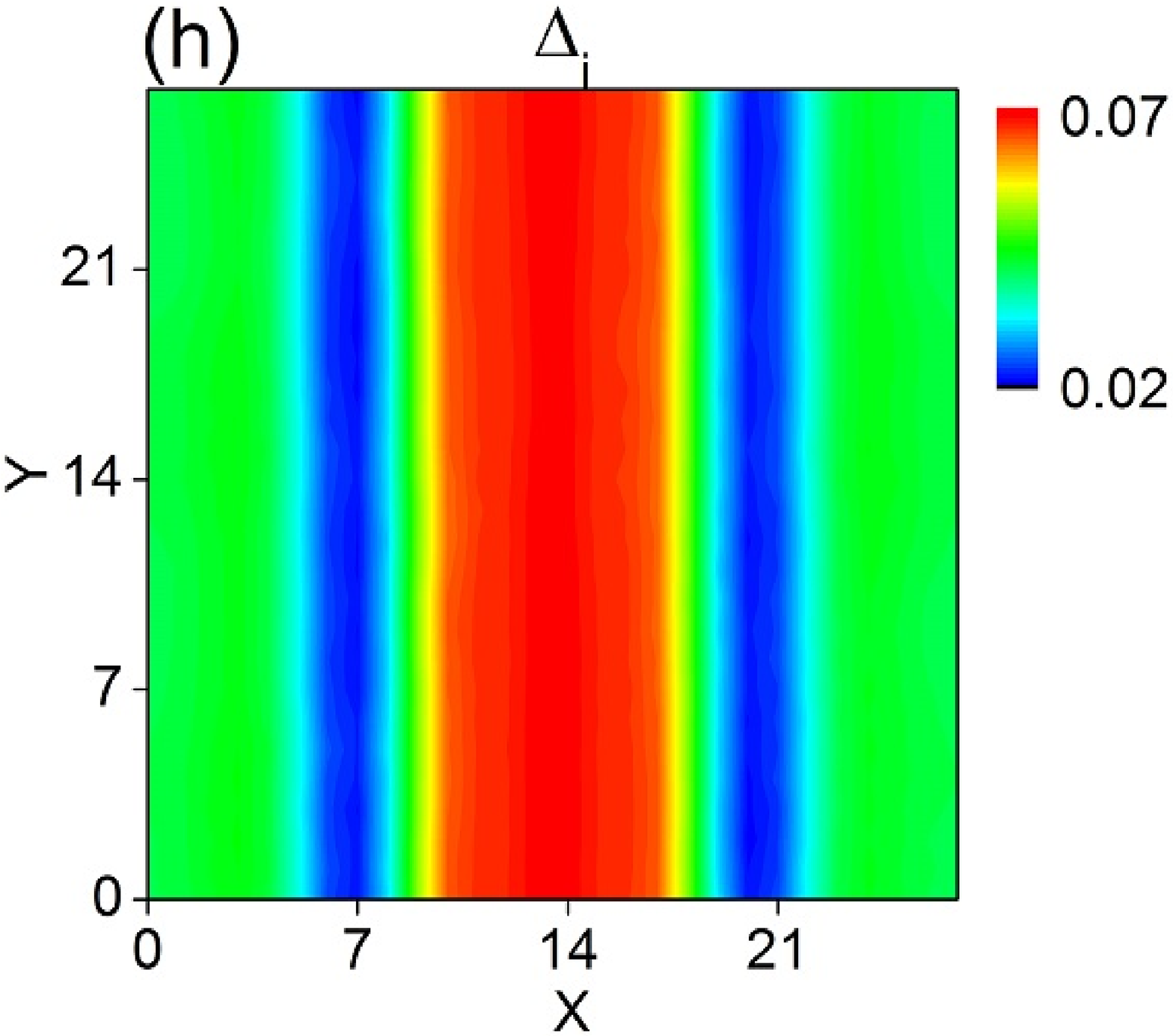}
\includegraphics[width=2.0in] {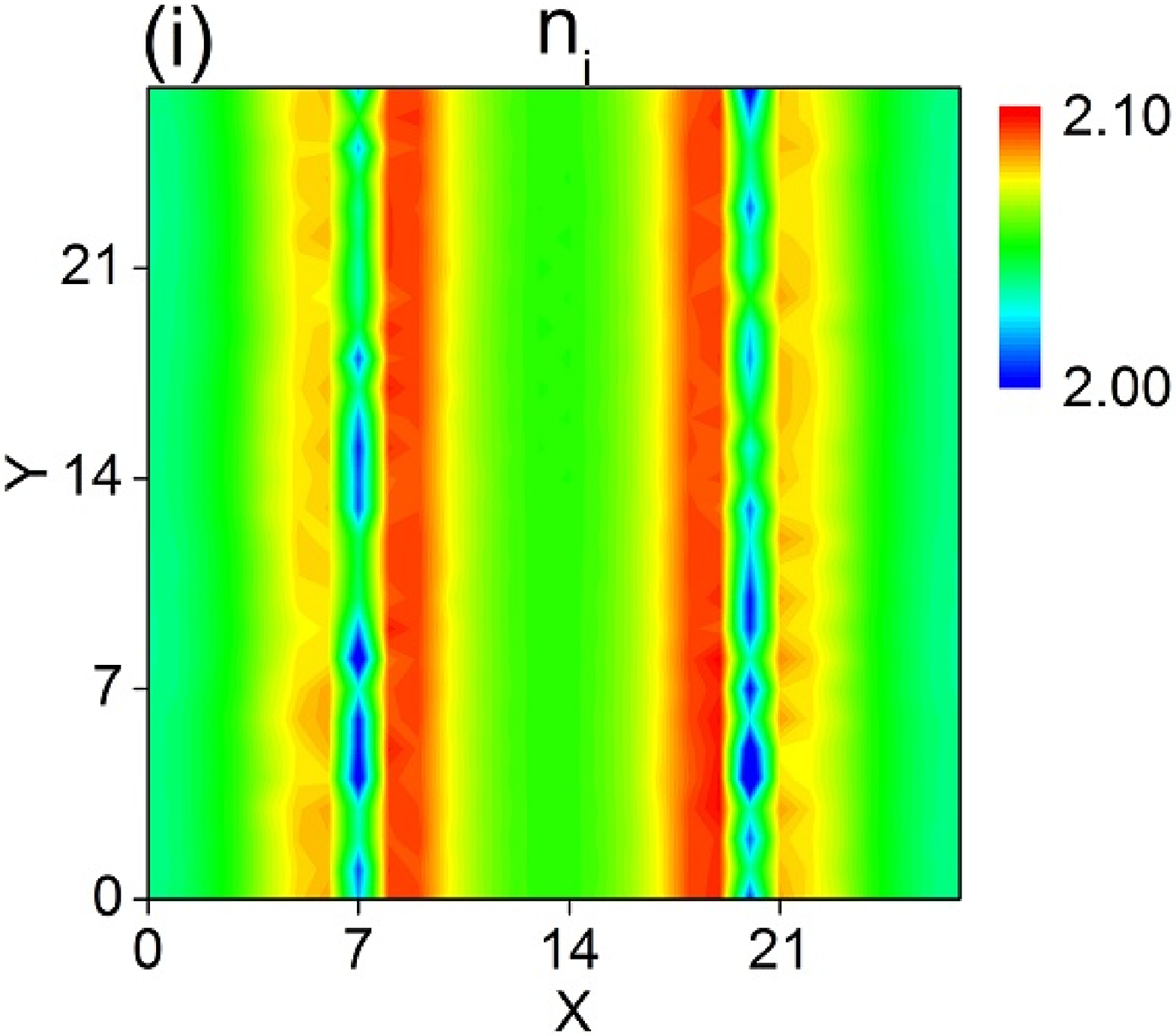}
\caption{The spatial profiles of magnetic order, superconducting
order and charge density for different mismatch scattering potential
$V_s=w$, where $w$ is randomly chosen from the range (-0.5, 0.5)
(($a$), ($b$) and ($c$)), range (-1.0, 1.0) (($d$), ($e$) and ($f$))
and range (-2.0, 2.0) (($g$), ($h$) and ($i$)).} \label{Figmm}
\end{figure}

All of our numerical results for the two types of TBs are calculated
in the under-doped region ($x=0.04$). For the parent compound with
$x=0$, the magnetic domain-wall structures still remain but there is
no SC over the whole sample.

\section{Twin-Boundary With Lattice Mismatch}
In the previous section we assume the lattices are well matched at
the TBs. In the case that the lattice on both sides of the TB are
not well matched, such as the case B in Section 4.2, the electrons
could subject to strong disordered scattering along the TBs. To take
this effect into consideration, we add an impurity-like potential of
random strength to each of the Fe ions along the TBs. The
Hamiltonian of this part is
$H_{scat}=\sum_{{s}\mu\sigma}V_{s}c^{\dagger}_{{s}\mu\sigma}
c_{{s}\mu\sigma}$, where $V_{s}$ is the impurity potential at the
${s-th}$ Fe site along the TBs.

Figure \ref{Figmm} shows the magnetic order, SC order and charge
density for a random scattering potential $V_s=w$ along the TBs. The
strengths of $w$ are chosen to be randomly distributed within the
ranges $(-0.5,0.5)$, $(-1.0, 1.0)$ and $(-2.0, 2.0)$ respectively
for figures \ref{Figmm}($a$), ($b$), ($c$), figures
\ref{Figmm}($d$), ($e$), ($f$) and figures \ref{Figmm}($g$), ($h$),
($i$). The results are averaged over 100 configurations. In figures
\ref{Figmm}($a$), ($b$) and ($c$) with weak scattering potential
$w=(-0.5, 0.5)$, the periodic modulations in the magnetic, SC and
the charge density orders which appearing in figure \ref{Figb} are
now smoothed by the weak disorder. The magnetic order is somewhat
suppressed along the TBs by he disorder, the SC gets enhanced near
the DW regions. If the random potential is increased to $w$=$(-1.0,
1.0)$, the results are shown in figures \ref{Figmm}($d$), ($e$) and
($f$). The SC order (see figure \ref{Figmm}($e$)) is now becoming
weakened as compared to that in figure \ref{Figmm}($b$). When the
random potential gets strong, such as $w=(-2.0, 2.0)$, the magnetic,
SC and charge density orders are exhibited in figures
\ref{Figmm}($g$), ($h$) and ($i$). In this case the SC is suppressed
on the magnetic DWs. Similar results could also be obtained in cases
A, C and D if the strength of mismatch across the TBs varying from
weak to strong. So far the observation of type-1 TB oriented
$90^\circ$ from the x-axis (as shown in case B) has not been
reported in the literatures, but an unpublished work of Pan
\cite{Pan} detected a rugged-shaped such kind of TB by STM
experiments in CaFe$_2$As$_2$, and indicates that the SC order is
greatly suppressed there. This is consistent with the our
theoretical study in the present section.

There also existed studies \cite{Prozorov, Li} in a magnetic field
in which the magnetic vortices are pinned at the TBs in the
subsection $4.1$. This result indicates that the SC is suppressed
along the TBs, and appears to be contrasting to our conclusion. But
one could not rule out the possibility that the magnetic vortex
might be pinned by the defects (impurities and vacancies) along the
TB, the SC would thus be suppressed there. In a very recent STM
experiment for FeSe \cite{Song}, similar type-1 TBs oriented
$45^\circ$ from the x-axis were detected but the SC is suppressed on
the TBs. This can be explained either by the fact that the electron
structure of FeSe is different from that of Ba(Ca)(FeAs)$_2$, or
there exist strong disordered scatterings along the TBs.

\section{Comparing With NMR Experiments}

\begin{figure}[t]
\centering
\includegraphics[width=2.5in] {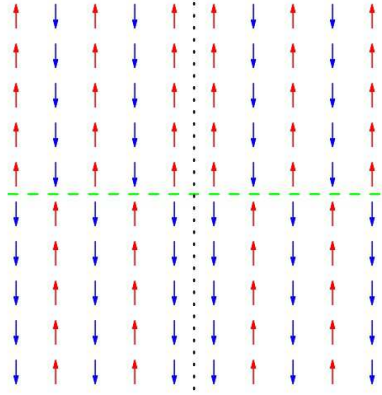}
\caption{Two types of magnetic domain walls observed in \cite{Xiao}.
The green-dashed line representing the antiphase DW while the
black-dotted line representing the DW across which the nearest
neighboring spins of the Fe atoms are ferromagnetically oriented.
The arrow here indicating the direction of the spin or the magnetic
moment of an Fe atom.} \label{figDW}
\end{figure}

\begin{figure}[t]
\centering
\includegraphics[width=2.0in] {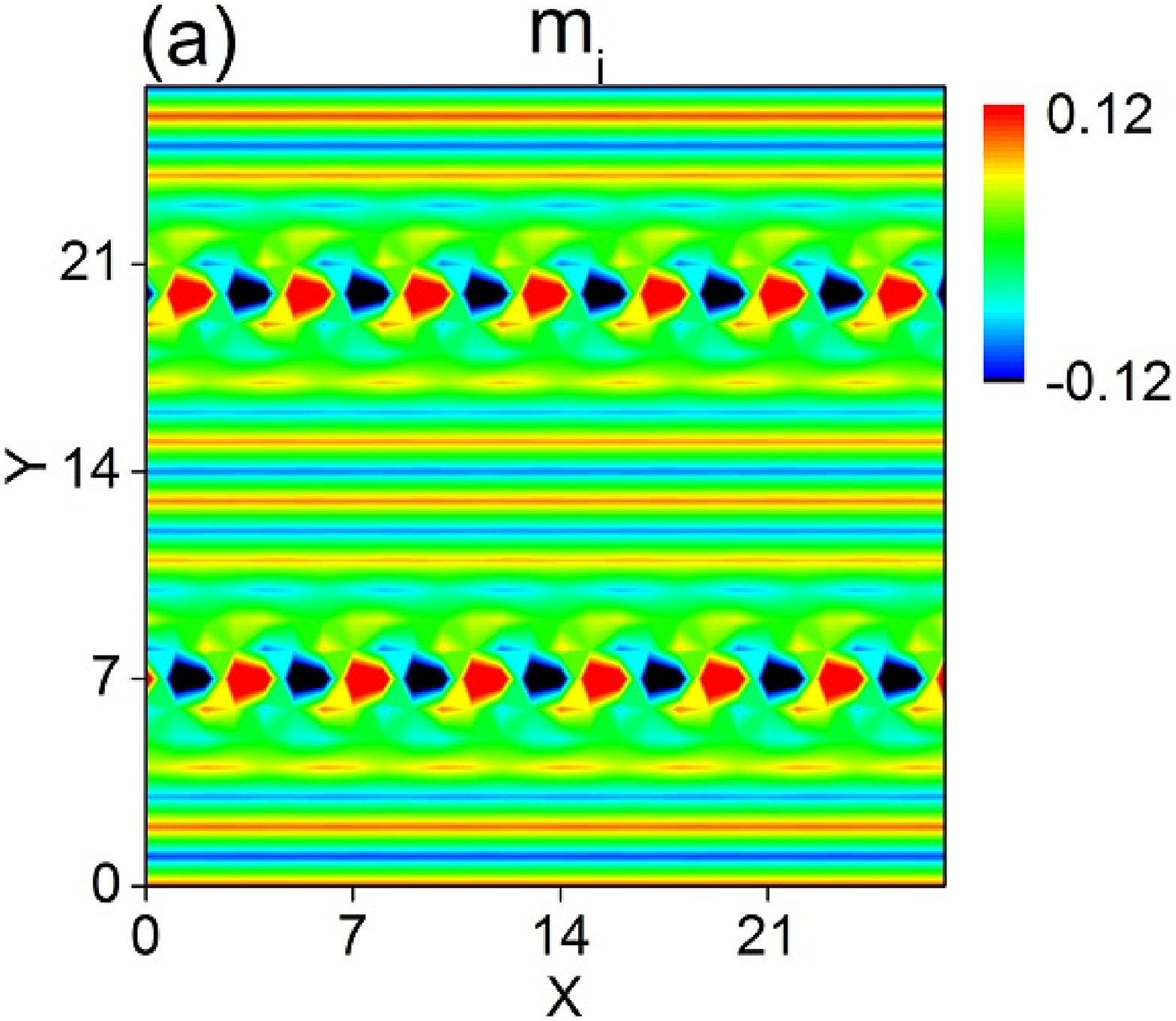}
\includegraphics[width=2.0in] {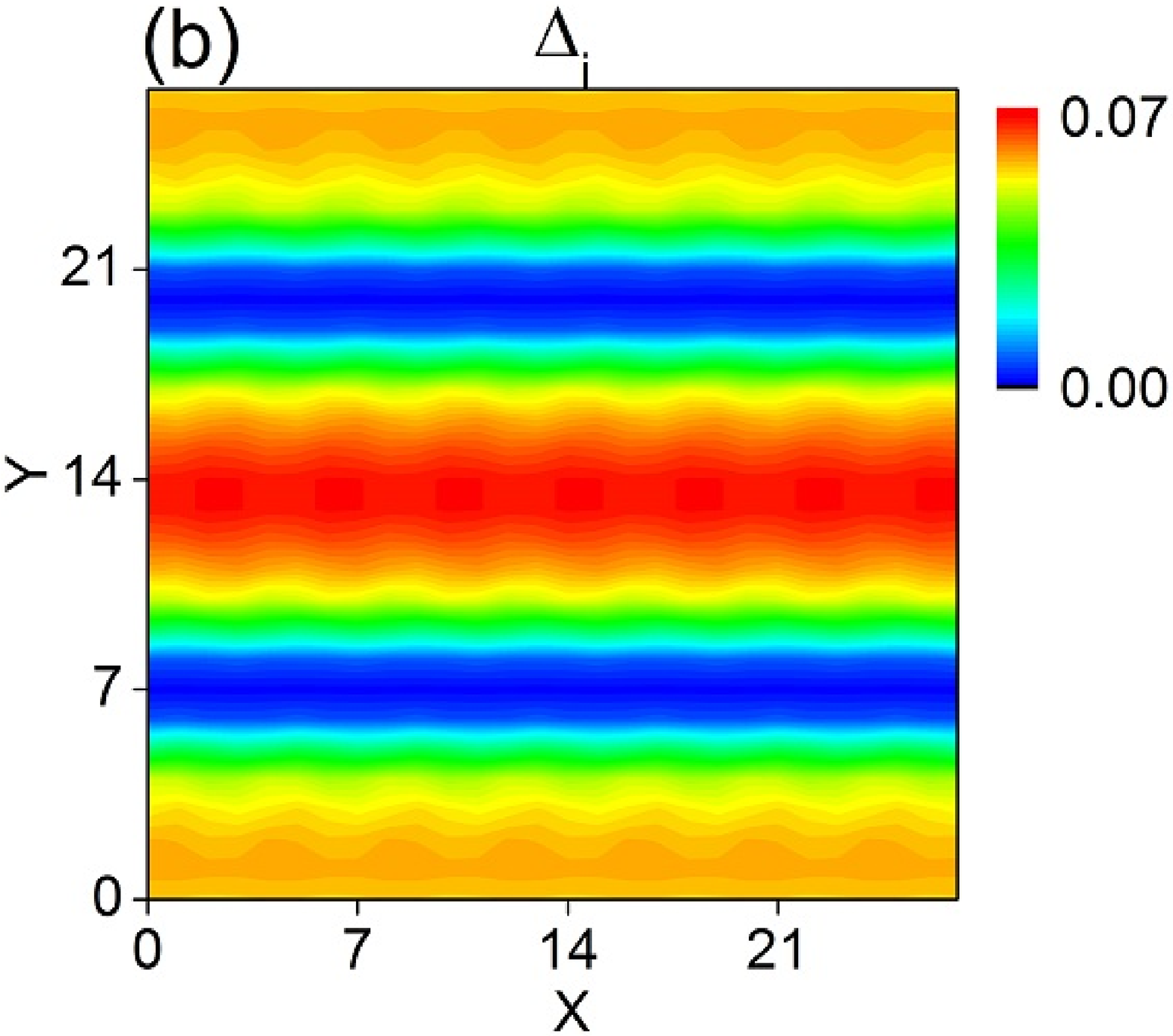}
\includegraphics[width=2.0in] {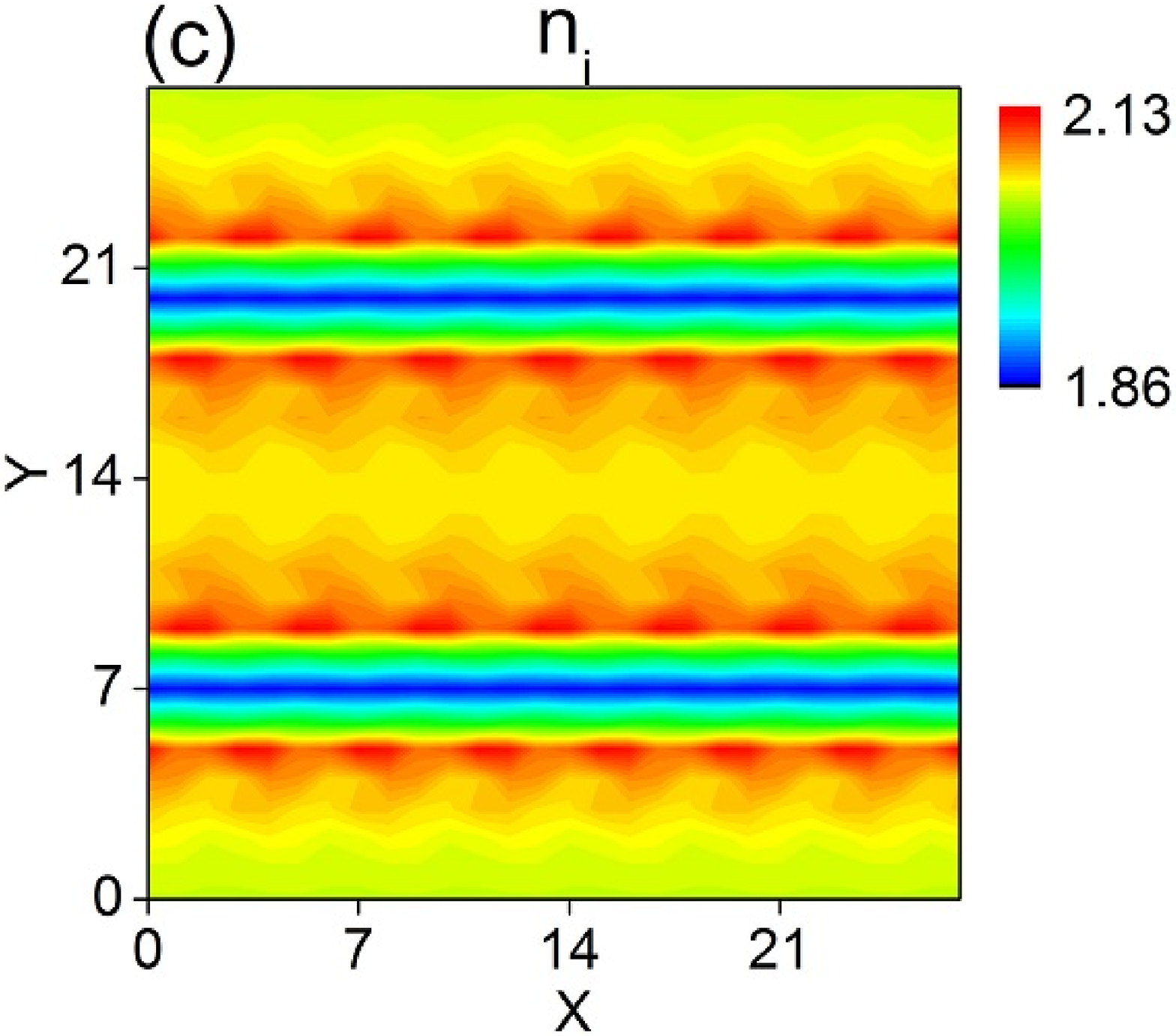}
\caption{Spatial profiles of ($a$) the magnetic order, ($b$) the
superconducting order, and ($c$) the charge density order are
presented for a type-2 TB oriented along x-axis.} \label{FigAnti}
\end{figure}

Recently there were indirect measurements of the magnetic structures
in a very underdoped or undoped Ca(FeAs)$_2$ sample by the NMR
experiments \cite{Xiao}. The authors not only observed the magnetic
DW along the $45^\circ$-oriented type-1 TB as demonstrated in
figures \ref{Figa0}($a$) and \ref{Figa}($a$), but also detected the
antiphase magnetic DWs as indicated by the green-dashed and the
black-dotted lines as shown in figure \ref{figDW}. The SC is found
to be enhanced on these DWs. It is still unclear what type of TBs or
defect lines are able to generate such anti-phase DWs. Here we wish
to point out that a similar aniphase DW like the black-dotted line
in figure \ref{figDW} could be generated by a type-2 TB (see figure
\ref{Figc0}), but oriented along x-axis. The results for the
magnetic, SC and charge density orders are shown in figures
\ref{FigAnti}($a$), ($b$) and ($c$) respectively. The magnetic DWs
are pinned at the TBs (see figure \ref{FigAnti}($a$)) with a
spin-density wave of a period $4a$ existing along the DW. It is
straightforward to see that the phases of the magnetic order above
and below the DWs (or the TBs) in figure \ref{FigAnti}($a$) differ
from those in figures \ref{Figc}($a$) by $180^\circ$. Since the
charge density (see figure \ref{FigAnti}($c$)) along the DWs
corresponds to that of lightly hole doped case, the SC is suppressed
(see figure \ref{FigAnti}($b$)). Although figure \ref{FigAnti}($a$)
gives rise to antiphase DWs, the SC is suppressed on them. This is
not completely in agreement with experiments \cite{Xiao}.

\begin{figure}[t]
\centering
\includegraphics[width=3.1in] {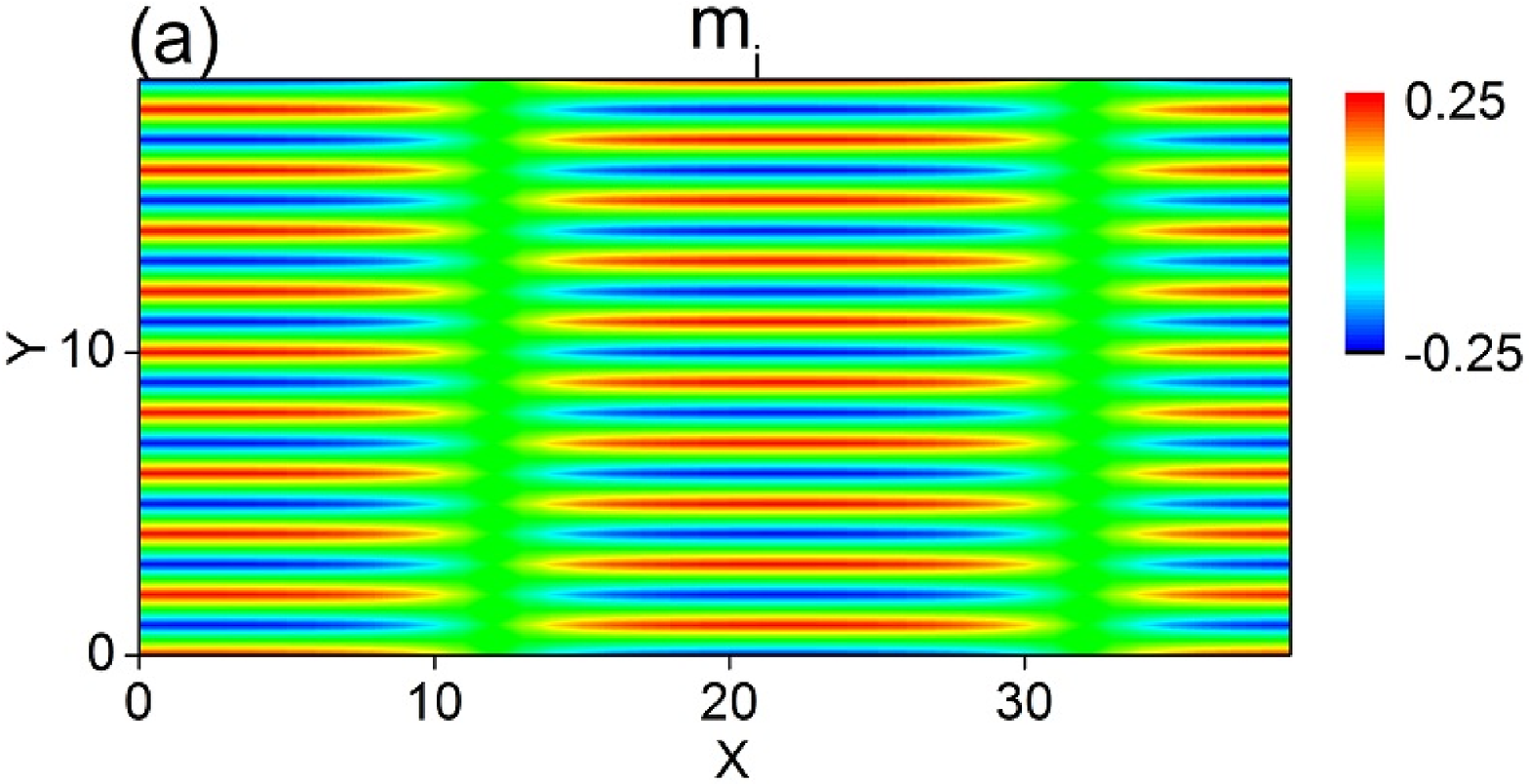}
\includegraphics[width=3.1in] {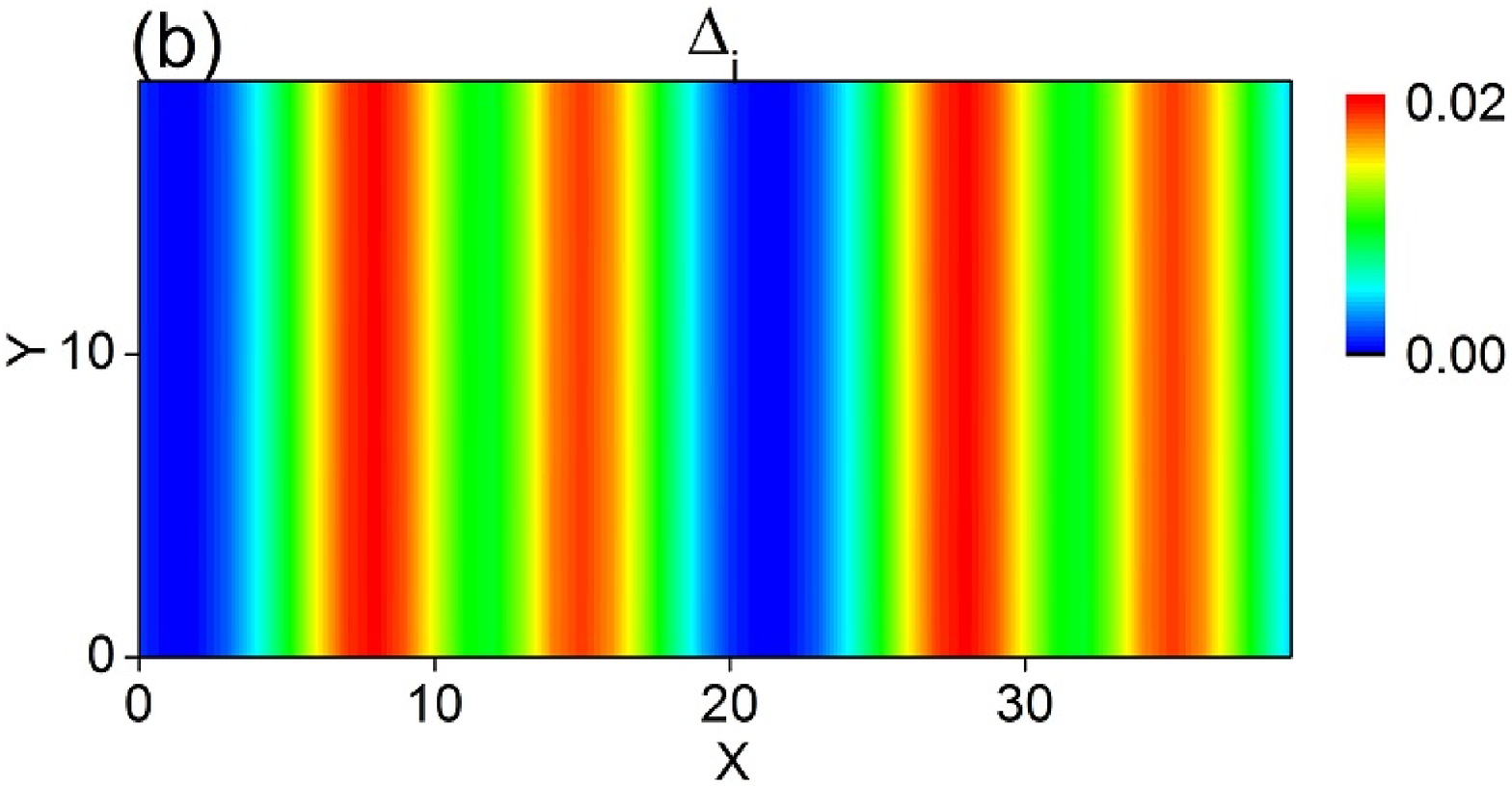}
\includegraphics[width=3.1in] {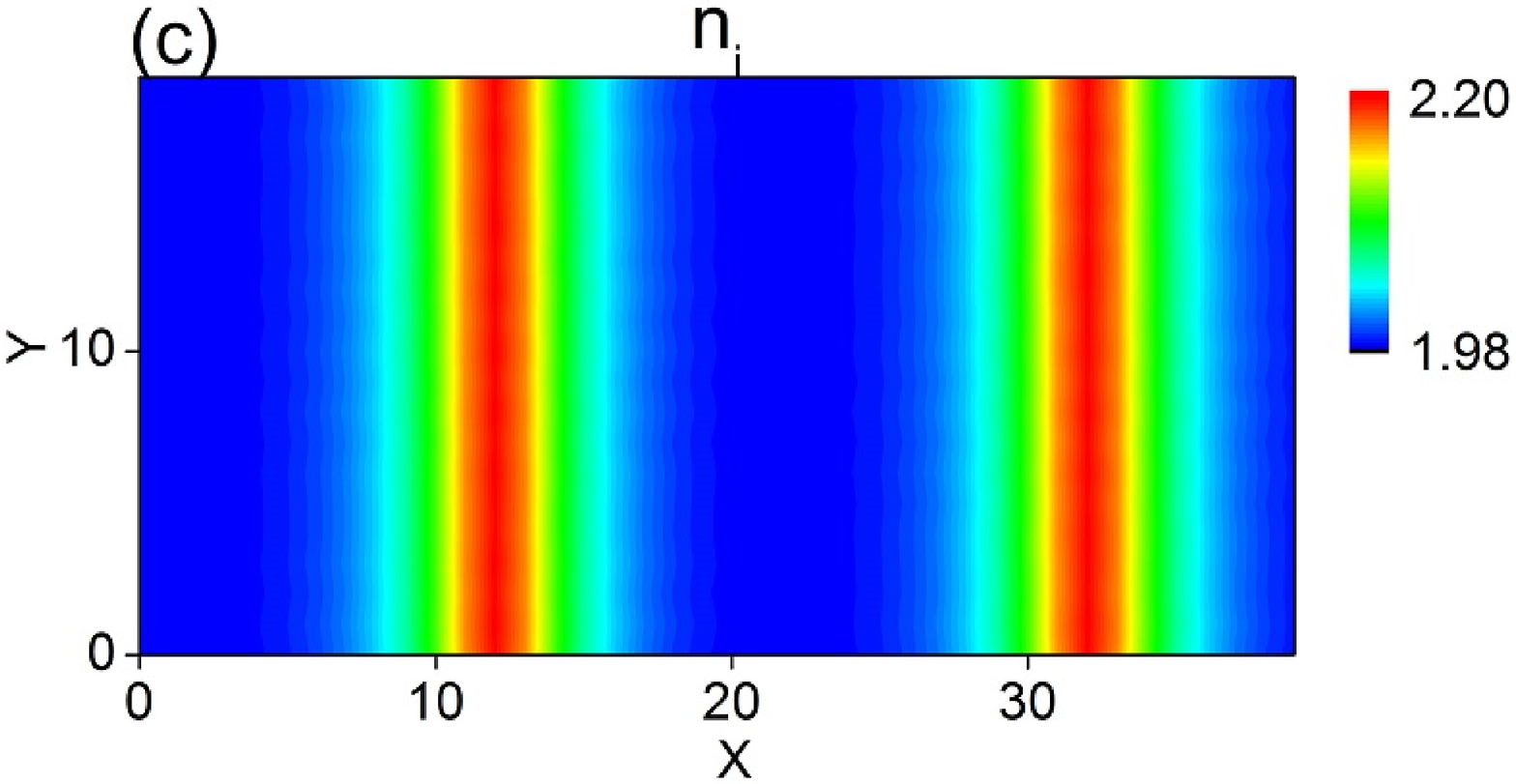}
\caption{Spatial profiles of ($a$) the magnetic order, ($b$) the
superconducting order, and ($c$) the charge density order in a
$20\times40$ lattice.} \label{figDWabc}
\end{figure}

Since the formation of DWs may not need TBs, they could be induced
by magnetic interactions \cite{Huang}. It is useful to note that in
the work of \cite{Huang}, only the magnetic DW in figure
\ref{Figa}($a$) appears to be stable at very low doping and there
exist no other types of DWs. In order to understand the experimental
observations, we try to numerically simulate the magnetic DWs by
choosing a stronger U. What we found is that the magnetic structure
shown in figure \ref{Figa}($a$) without the TBs \cite{Huang} is
always stable, even when the onsite Coulomb interaction $U$ is
increased to a reasonable magnitude. As $U$ is increased from 3.8 to
5.5, similar DWs to that indicated by the green-dashed line in
figure \ref{figDW} could also be generated (see figure
\ref{figDWabc}($a$)), in addition to the DWs in figure
\ref{Figa}($a$). The magnetic DW is defined where the magnetic order
parameter is suppressed the most. On the opposite sides of the DWs,
the magnetic moments of the Fe atoms point to opposite directions.
This should correspond to the antiphase DWs predicted in
\cite{Mazin}. However, the magnetic configuration in figure
\ref{figDWabc}($a$) appears to have slightly higher energy than that
in figure \ref{Figa}($a$), and thus it should be regarded as a
meta-stable or local stable. This type of DWs may become detectable
in experiments under proper local condition of the sample. In figure
\ref{figDWabc}($b$), the spatial distribution of the SC order
parameter is presented, and it is shown that the SC order is
suppressed at the middle of the magnetic domains where the charge
density corresponding to that of slightly hole doped case (see
figure \ref{figDWabc}($c$)), but enhanced near the boundary between
the magnetic domain or the edges of the DW where the charge density
corresponding to the optimal electron-doped case. On the DWs, the
charge density is in the overdoped region where the magnetic order
is completely suppressed. Here, it needs to point out that the width
of our numerically obtained DWs as showing in figures
\ref{FigAnti}($a$)and \ref{figDWabc}($a$) span several lattice
constants. While in the experiments \cite{Xiao} (or see figure
\ref{figDW}), the width of the DW covers only one lattice constant.
This should not be the physical picture, because it would cost large
exchange energy at the DW, as one-line of spins being flipped among
the two adjacent spin lines close to the DW. We also spent a lot of
effort trying to numerically simulate the magnetic DW indicated by
the black-dotted line as shown in figure \ref{figDW} using only $U$
but no TBs. This task so far has not been successful, and it implies
that the cost of energy in creating such a DW is high, because one
has to flip the spins of the Fe atoms on the right hand side of the
DW in order to make the nearest neighboring spins across the DW
ferromagnetically oriented. Since the NMR experiments \cite{Xiao}
have indirectly detected all the three magnetic DWs discussed above,
which could indicate that their sample may not be homogeneous so
that it could accommodate all these DWs of different characteristics
in different parts of the sample. It is also interesting to note
that the antiphase DWs so far have not been confirmed by other more
direct measurements.

In summary for this section, although the results obtained from the
present work seem able to qualitatively reproduce the the antiphase
DWs or boundaries as observed by experiments, the positions of the
enhanced SC are not at the DWs, instead they are near the edges of
the DWs. We either have to find the proper TBs or the defect lines
for generating exactly the kind of antiphase DWs with the SC pinned
at the DWs as observed by the NMR experiments, or the experimental
measurements have to be reinterpreted.

\section{Conclusions}
Based on the BdG equations and the mean-field approach, we study the
effects of TBs on the complex competition between the magnetism and
superconductivity in slightly electron-doped Ba(Ca)(FeAs)$_2$
compounds, particularly, to examine whether the SC order would be
enhanced at the TBs. There are three points that need to be
emphasized here. First, the formation and the location of the DWs
strongly depend on the nature of TBs. For the four kinds of TBs
studied in Section 4, the DWs in cases A, B and D are found to be
pinned at the TBs, while in case C the DWs are separated from the
TBs. Intuitively, the electrons are subject to additional potentials
induced by the TBs. For cases A and B, these potentials are
negative, and thus the electron density is slightly enhanced on the
TBs. This causes a higher electron-doping level and the SC get
enhanced near the TBs than on other sites. The magnetic order near
the TBs in this case becomes weakened. Additionally, the $90^\circ$
lattice rotation across the TB splits the degeneracy between
different orientated $2\times1$ AF states. Therefore $90^\circ$
orientated AF states are favored on different sides of the TBs, and
the DWs are naturally pinned at the TBs. For cases C and D, the
additional potentials for the electrons appear to be positive, and
thus the electron density is decreased to such a value that the
effective doping level corresponds to that of hole-doping along the
TBs. For case C, the charge density along the TB is slightly hole
doped while the magnetic order is still quite strong. As a result,
the SC near the TB becomes quite suppressed. As a result, the DW is
located between two neighboring TBs and the SC gets much enhanced
there. For case D, the charge density near the TB corresponds to the
over-hole doped case. The strengths of the magnetic order and the SC
are greatly reduced there. Thus the DW is pinned near the TB. This
naturally explains why the magnetic order is enhanced along the TBs
in case C and becomes rather weak in case D. Moreover, the
reflection symmetry with respect to the TBs is preserved in case C
and broken in case D. This makes it more difficult to form a
symmetric magnetic order near the TBs in case D than in case C.

Second, the formation of the DWs implies that the magnetism is
Inhomogeneous, which is accompanied by the non-uniform distributions
of the SC and the charge density orders. In cases of A, B and C, the
SC is enhanced in the regions near the DWs where The magnetic order
is suppressed. The reason for this is that the electron densities
are enhanced to the optimal doped level around the DWs. However in
case D, on the DWs where the  magnetic order is suppressed, the
carrier density is close to the over(hole)-doping level, and this is
also unfavorable to SC. As a result, SC coexists with the magnetism
in the middle of magnetic domains. In summary, the SC along the TBs
gets enhanced in cases A and B, while it is much suppressed in cases
C and D. The predictions for cases C and D could be tested by
measuring the superfluid density on the TBs using SQIDM
\cite{Kalisky}.

Finally, we point out that our results on the type-1 TB oriented
$45^\circ$ from the x-axis (see figure \ref{Figa}) are in good
agreement with experiments \cite{Chuang,Kalisky}. The two types of
antiphase DWs or boundaries as detected indirectly by NMR
experiments \cite{Xiao} could also somewhat be generated
respectively by a type-1 TB oriented along x-axis and a larger
onsite Coulomb interaction $U$. The SC is found to be enhanced near
the edges of the DWs, but not at the DWs as observed in the
experiments.

\ack
 This work was supported by the Texas Center for
Superconductivity at the University of Houston, by the Robert A.
Welch Foundation under Grant No. E-1146, and by the NSF through
grants DMR-0908286 and DMR-1206839.


\section*{References}
\begin{thebibliography}{99}
\bibitem{Kamihara} Kamihara Y, Watanabe T, Hirano M and Hosono H 2008 {\it J. Am. Chem. Soc.} {\bf 130} 3296
\bibitem{Ren} Ren Z A \etal 2008 {\it Chin. Phys. Lett.} {\bf 25} 2215
\bibitem{Chen} Chen X H, Wu T, Wu G, Liu R H, Chen H and Fang D F 2008 {\it Nature} {\bf 453} 761
\bibitem{Cruz} de la Cruz C \etal 2008 {\it Nature} {\bf 453} 899
\bibitem{Chen08} Chen G F, Li Z, Wu D, Li G, Hu W Z, Dong J, Zheng P, Luo J L and Wang N L 2008 \PRL {\bf 100} 247002
\bibitem{Laplace} Laplace Y, Bobroff J, Rullier-Albenque F, Colson D and Forget A 2009 \PR B {\bf 80} 140501
\bibitem{Julien} Julien M -H, Mayaffre H, Horvatic M, Berthier C, Zhang X D, Wu W, Chen G F, Wang N L and Luo J L 2009 {\it Europhys. Lett.} {\bf 87} 37001
\bibitem{Mazin} Mazin I I and Johannes M D 2009 {\it Nature Phys.} {\bf 5} 141
\bibitem{Gorkov} Gor'kov L P and Teitel'baum G B 2010 {\it arXiv}:{\bf 1001}.4641
\bibitem{Chuang} Chuang T M, Allan M P, Lee J, Xie Y, Ni N, Bud'ko S L, Boebinger G S, Canfield P C and Davis J C 2010 {\it Science} {\bf 327} 181
\bibitem{Kalisky} Kalisky B, Kirtley J R, Analytis J G, Chu J -H, Vailionis A, Fisher I R and Moler K A 2010 \PR B {\bf 81} 184513
\bibitem{Huang} Huang H, Zhang D, Zhou T and Ting C S 2011 \PR B {\bf 83} 134517
\bibitem{Zhang} Zhang D 2009 \PRL {\bf 103} 186402
\bibitem{Zhou} Zhou T, Zhang D and Ting C S 2010 \PR B {\bf 81} 052506
\bibitem{Wang} Wang X F, Wu T, Wu G, Liu R H, Chen H, Xie Y L and Chen X H 2009 \NJP {\bf 11} 045003
\bibitem{Pratt} Pratt D K, Tian W, Kreyssig A, Zarestky J L, Nandi S, Ni N, Bud'ko S L, Canfield P C, Goldman A I and McQueeney R J 2009 \PRL {\bf 103} 087001
\bibitem{Christianson} Christianson A D, Lumsden M D, Nagler S E, MacDougall G J, McGuire M A, Sefat A S, Jin R, Sales B C and Mandrus D 2009 \PRL {\bf 103} 87002
\bibitem{Zhou11} Zhou T, Huang H, Gao Y, Zhu J and Ting C S 2011 \PR B {\bf 83} 214502
\bibitem{Terashima} Terashima K \etal 2009 {\it Proc. Natl. Acad. Sci. U. S. A.} {\bf 106} 7330
\bibitem{Sekiba} Sekiba Y \etal 2009 \NJP {\bf 11} 025020
\bibitem{Richard} Richard P \etal 2010 \PRL {\bf 104} 137001
\bibitem{Huynh} Huynh K K, Tanabe Y and Tanigaki K 2011 \PRL {\bf 106} 217004
\bibitem{Shan} Shan L \etal 2011 {\it Nature Physics} {\bf 7} 325
\bibitem{Gao} Gao Y, Huang H, Chen C, Ting C S and Su W -P 2011 \PRL {\bf 106} 027004
\bibitem{Huang1} Huang H, Gao Y, Zhu J -X and Ting C S 2012 \PRL {\bf 109} 187007
\bibitem{Yong} Yong J, Lee S, Jiang J, Bark C W, Weiss J D, Hellstrom E E, Larbalestier D C, Eom C B and Lemberger T R 2011 \PR B {\bf 83} 104510
\bibitem{Xiao} Xiao H \etal 2012 \PR B {\bf 85} 024530
\bibitem{Seo} Seo K, Bernevig B A and Hu J 2008 \PRL {\bf 101} 206404
\bibitem{Mazin08} Mazin I I, Singh D J, Johannes M D and Du M H 2008 \PRL {\bf 101} 057003
\bibitem{Kim} Kim M G, Fernandes R M, Kreyssig A, Kim J W, Thaler A, Bud'ko S L, Canfield P C, McQueeney R J, Schmalian J and Goldman A I 2011 \PR B {\bf 83} 134522
\bibitem{Kalisky1} Kalisky B, Kirtley J R, Analytis J G, Chu J -H, Fisher I R and Moler K A 2011 \PR B {\bf 83} 064511
\bibitem{Hu} Hu J and Hao N 2012 \PR X {\bf 2} 021009
\bibitem{Pan} Pan S H and Wang J H 2013 private communication
\bibitem{Prozorov} Prozorov R, Tanatar M A, Ni N, Kreyssig A, Nandi S, Bud'ko S L, Goldman A I and Canfield P C 2009 \PR B {\bf 80} 174517
\bibitem{Li} Li L J, Nishio T, Xu Z A and Moshchalkov V V 2011 \PR B {\bf 83} 224522
\bibitem{Song} Song C -L, Wang Y -L, Jiang Y -P, Wang L, He K, Chen X, Hoffman J E, Ma X -C and Xue Q -K 2012 \PRL {\bf 109} 137004
\end{thebibliography}
\end{document}